\documentclass[preprint,manuscript,11pt]{aastex}

\def\chandra{{\it Chandra\/}}

\def\genx{{\it Generation-X\/}}
\def\einstein{{\it Einstein\/}}
\def\hst{{\it {\it HST}\/}}

\def\rosat{{\it ROSAT\/}}

\def\xeus{{\it XEUS\/}}

\def\xmm{{\it XMM-Newton\/}}

\def\xray{\hbox{X-ray}}
\def\ecdfs{\hbox{E-CDF-S}}
\def\cdfs{\hbox{CDF-S}}
\def\cdfn{\hbox{CDF-N}}
\def\etal{{et\,al.\,}}

\def\ltsima{$\; \buildrel < \over \sim \;$}
\def\simlt{\lower.5ex\hbox{\ltsima}}
\def\gtsima{$\; \buildrel > \over \sim \;$}
\def\simgt{\lower.5ex\hbox{\gtsima}}
\def\kms{\ifmmode{~{\rm km~s^{-1}}}\else{~km s$^{-1}$}\fi}
\def\lsim{\lower0.3em\hbox{$\,\buildrel <\over\sim\,$}}
\def\gsim{\lower0.3em\hbox{$\,\buildrel >\over\sim\,$}}
\def\lbsol{$L_{B,\odot}$}
\def\msol{$M_\odot$}
\def\h2{H$_2$}
\def\flux{ergs~cm$^{-2}$~s$^{-1}$}
\def\xlum{ergs~s$^{-1}$}

\def\arcsec{\mbox{$^{\prime\prime}$}}
\def\arcmin{\mbox{$^\prime$}}

\def\Lx{$L_{\rm X}$}

\begin{document}

\shortauthors{LEHMER ET AL.}
\shorttitle{X-ray Evolution of Early-Type Galaxies}

%
\title{The X-ray Evolution of Early-Type Galaxies in the Extended \chandra\ Deep Field-South}
%

\author{
B.~D.~Lehmer,\altaffilmark{1}
W.~N.~Brandt,\altaffilmark{1}
D.~M.~Alexander,\altaffilmark{2}
E.~F.~Bell,\altaffilmark{3}
D.~H.~McIntosh,\altaffilmark{4}
F.~E.~Bauer,\altaffilmark{5}
G.~Hasinger,\altaffilmark{6}
V.~Mainieri,\altaffilmark{6}
T.~Miyaji,\altaffilmark{7}
D.~P.~Schneider,\altaffilmark{1}
\& A.~T.~Steffen\altaffilmark{1}
}
\altaffiltext{1}{Department of Astronomy \& Astrophysics, 525 Davey Lab,
The Pennsylvania State University, University Park, PA 16802, USA}
\altaffiltext{2}{Department of Physics, University of Durham, South Road, Durham, DH1 3LE, UK}
\altaffiltext{3}{Max-Planck-Institut f\"ur Astronomie, K\"onigstuhl 17, D-69117 Heidelberg, Germany}
\altaffiltext{4}{Astronomy Department, University of Massachusetts, 710 N. Pleasant St., Amherst, MA 01007, USA}
\altaffiltext{5}{Columbia Astrophysics Laboratory, Columbia University, Pupin Labortories, 550 W. 120th St., Rm 1418, New York, NY 10027, USA}
\altaffiltext{6}{Max-Planck-Institut f\"ur extraterrestrische Physik, Giessenbachstrasse, D-85748 Garching b. M\"unchen, Germany}
\altaffiltext{7}{Department of Physics, Carnegie Mellon University, Pittsburgh, PA 15213, USA}

%
\begin{abstract}
%

We investigate the evolution over the last 6.3~Gyr of cosmic time (i.e., since
$z \approx 0.7$) of the average \xray\ properties of early-type galaxies within
the  Extended \chandra\ Deep Field-South (\hbox{E-CDF-S}).  Our early-type
galaxy sample includes 539 objects with red-sequence colors and S\'{e}rsic
indices larger than $n = 2.5$, which were selected jointly from the COMBO-17
(Classifying Objects by Medium-Band Observations in 17 Filters) and GEMS
(Galaxy Evolution from Morphologies and SEDs) surveys.  We utilize the deep
\chandra\ observations over the \hbox{E-CDF-S} and \xray\ stacking analyses to
constrain primarily the average \hbox{X-ray} emission from ``normal''
early-type galaxies (i.e., those that are not dominated by luminous active
galactic nuclei [AGNs]).  In our analyses, we study separately optically
luminous ($L_B \approx 10^{10-11}$~\lbsol) and faint ($L_B \approx
10^{9.3-10}$~\lbsol) galaxy samples, which we expect to have soft
(\hbox{0.5--2.0~keV}) \xray\ emission dominated by hot ($\sim$1~keV)
interstellar gas and low mass \xray\ binary (LMXB) populations, respectively.
We detect individually 49 ($\approx$9\%) of our galaxies in the \xray\ band,
and classify these sources as either normal early-type galaxies (17 galaxies)
or AGN candidates (32 galaxies).  The AGN fraction of our optically luminous
samples evolves with redshift in a manner consistent with the $(1+z)^3$
evolution observed in other investigations of \xray-selected AGNs.  After
removing potential AGNs from our samples, we find that the
\hbox{\xray--to--$B$-band} mean luminosity ratio ($L_{\rm X}/L_B$) for
optically luminous early-type galaxies does not evolve significantly over the
redshift range \hbox{$z \approx$~0.0--0.7}.  This lack of \xray\ evolution
implies a general balance between the heating and cooling of the hot
interstellar gas.  If transient AGN activity is largely responsible for
maintaining this balance, then we infer that mechanical power must be
dominating the feedback out to $z \approx 0.7$.  Furthermore, in this scenario
the average mechanical AGN power must remain roughly constant over the last
half of cosmic time despite significant evolution in the average AGN radiative
luminosity.  For our optically faint early-type galaxies, we find suggestive
evidence that $L_{\rm X}/L_B$ increases with redshift over the range \hbox{$z
\approx$~0.0--0.5}; however, due to limited statistical constraints on both the
local $L_{\rm X}/L_B$ ratio and the AGN contamination of our samples, we
consider this result to be marginal.  

%
\end{abstract}
%


%
\section{Introduction}
%

Distant (\hbox{$z \approx$~0.1--3}), isolated early-type galaxies have been
studied extensively at optical and near-IR wavelengths.  The optical
luminosities of early-type galaxies with rest-frame red-sequence colors have
faded by $\approx$1~magnitude since $z=1$ (e.g., Bell \etal 2004a; McIntosh
\etal 2005; Treu \etal 2005, 2006; van~der~Wel \etal 2005).  Furthermore,
little star-formation activity is observed within these galaxies out to at
least $z \approx 0.7$ (e.g., Bell \etal 2005) indicating that this observed
evolution is consistent with the passive aging of old stars.  However, the
luminosity density of red-sequence galaxies has been observed to remain
approximately constant over \hbox{$z \approx$~0.1--0.7} (e.g., Bell \etal
2004a; Faber \etal 2006), and since these galaxies appear to be fading
passively with cosmic time, this suggests that an important fraction
($\approx$1/2) of the current $z=0$ stellar-mass content of the early-type
galaxy population has emerged since $z = 1$.  One explanation for these
observations is that relatively young, lower-mass galaxies continuously
populate the red-sequence (e.g., van~der~Wel \etal 2004; Thomas \etal 2005)
through either a downsizing mass-assembly scenario (e.g., Cowie \etal 1996;
Bundy \etal 2005; Cimatti \etal 2006; De~Lucia \etal 2006; Lee \etal 2006)
and/or a scenario where lower-mass early-type galaxies have different merging
histories than their higher-mass counterparts (e.g., Khochfar \& Burkert~2003).
At $z \simgt 1$, distant red galaxies (DRGs; selected using $J-K$ colors) and
extremely red objects (EROs; $R-K$ or $I-K$ selected), some of which are likely
progenitors to early-type galaxies, appear to have large star-formation rates
and inferred dynamical masses similar to those observed for massive local
ellipticals (e.g., F{\"o}rster--Schreiber \etal 2004; McCarthy 2004; van~Dokkum
\etal 2004; Papovich \etal 2006).  These analyses suggest that the majority of
the stellar mass within many massive early-type galaxies formed at $z \simgt
1.5$, possibly during a dust-obscured far-infrared-luminous phase (e.g.,
Chapman \etal 2005; Swinbank \etal 2006).

\subsection{X-ray Properties of Early-Type Galaxies}

X-ray investigations of local early-type galaxies have revealed that the total
\xray\ luminosity is correlated with optical luminosity (e.g., Trinchieri \&
Fabbiano 1985; Canizares \etal 1987; Brown \& Bregman 1998; O'Sullivan \etal
2001; Ellis \& O'Sullivan 2006).  \xray\ emission from local early-type
galaxies originates primarily from a combination of hot interstellar gas and
low mass \xray\ binaries (LMXBs), and due to their relatively low
star-formation rates (SFRs) per unit stellar mass (typically SFR/$M_* \approx
10^{-3}$--$10^{-1}$~Gyr$^{-1}$ for $z \approx$~0.0--0.7 galaxies with $M_*
\simgt 10^{10}$~\msol; e.g., Feulner \etal 2006), these galaxies are expected
to have negligible contributions from high mass \xray\ binaries (HMXBs; e.g.,
Grimm \etal 2002).  Two fairly distinct classes of early-type galaxies, which
can be divided by optical luminosity at \hbox{$L_{B,\rm crit} \approx 10^{10}$
\lbsol}, have been identified.  Early-type galaxies with $L_B \simgt L_{B, \rm
crit}$ are referred to as ``optically luminous'' and those with $L_B \simlt
L_{B, \rm crit}$ are referred to as ``optically faint.''  

Optically luminous early-type galaxies in the local universe are generally
observed to have \xray\ emission powered largely by hot (\hbox{$kT
\approx$~0.3--1~keV}) interstellar gas at soft (\hbox{0.5--2.0~keV}) \xray\
energies and LMXBs at hard (\hbox{2--10~keV}) \xray\ energies (e.g., Matsumoto
\etal 1997).  Figure~1a illustrates the \xray\ spectrum of NGC~1600 (Sivakoff
\etal 2004), a representative optically luminous early-type galaxy; the hot-gas
and LMXB components have been separated for illustrative purposes.  The \xray\
luminosities for optically luminous early-type galaxies are correlated with
their optical luminosities following \Lx~$\propto L_B^2$.  Stellar winds
generate the gas and high-velocity interactions and Type~Ia supernovae heat it
to \xray\ temperatures (e.g., Sarazin 1997).  Galaxies with larger stellar
masses and deeper gravitational potential wells generally have higher stellar
velocity dispersions and thus provide more significant heating of the gas.
Therefore, the \hbox{\Lx--$L_B$} relation is manifested as a result of
proportionalities between the stellar mass in a particular galaxy (traced by
$L_B$) and the stellar velocity dispersion of that galaxy (traced by \Lx; e.g.,
Mahdavi \& Geller 2001).

Optically faint early-type galaxies in the local universe have soft \xray\
emission characterized by a power-law (\hbox{$\Gamma \approx 1.8$}) component
from LMXBs and a mild contribution from hot interstellar gas (i.e., a thermal
plasma with $kT \sim 0.1$~keV).  At hard \xray\ energies, the emission is
completely dominated by LMXBs.  The total \xray\ luminosities of optically
faint early-type galaxies are observed to be linearly correlated with their
optical luminosities (i.e., \Lx~$\propto L_B$).  This linear correlation is
thought to arise largely as a result of the number of LMXBs in a given galaxy
being proportional to the stellar mass within that galaxy (e.g., Gilfanov
2004).  Due to their low luminosities, few optically faint early-type galaxies
have been studied in detail in the \xray\ band, and the presently available
data for these galaxies are plagued by large scatter due to small numbers of
LMXBs per galaxy and varying contributions from hot interstellar gas (e.g.,
David \etal 2006).  Therefore, much of what is known about LMXBs in these
galaxies comes from studying optically luminous early-type galaxies that have
little contamination from hot interstellar gas (i.e., \xray\ faint early-type
galaxies).  Figure~1b shows the \xray\ spectrum of NGC~4697 (e.g., Sarazin
\etal 2001), an \xray\ faint (yet optically luminous) early-type galaxy, which
is expected to be representative of the typical optically faint early-type
galaxy \xray\ spectrum.   

\subsection{Physical Properties of the X-ray Emitting Components}

The two early-type galaxy classes discussed above (optically luminous and
faint) generally differ in their relative spectral contributions from hot
interstellar gas and LMXBs.  An important component of the \xray\ emission from
early-type galaxies, most notably in optically luminous sources, is the hot
interstellar gas.  Detailed observations of the gas have shown that it radiates
strongly but does not appear to be cooling as expected (e.g., Xu \etal 2002;
$\S$~7.1 of Mathews \& Brighenti 2003 and references therein; Tamura \etal
2003; Bregman \etal 2005).  The observed gas temperatures and densities of the
central regions of optically luminous early-type galaxies (e.g., Fukazawa \etal
2006) imply short radiative cooling times of $\approx$$10^8$~yr on average.
Simple cooling-flow models, which include heating from stellar winds and
Type~Ia supernovae, vastly overpredict the amount of cooled gas observed in the
central regions of these galaxies (see, e.g., $\S$~8 of Mathews \& Brighenti
2003); this suggests some additional form of heating must be present.  Attempts
to mitigate this problem by including additional heating components in these
models, such as transient heating from active galactic nuclei (AGNs), have been
unsuccessful in reproducing the observed properties of the gas (e.g., the
radial profiles of electron temperature, density, and metallicity; see, e.g.,
Brighenti \& Mathews 2003).  Despite these modelling difficulties, both
observational and theoretical studies have suggested that energy feedback from
transient AGNs must be important (e.g., Binney \& Tabor 1995; Ciotti \&
Ostriker 1997, 2001; B{\^i}rzan \etal 2004; Best \etal 2005, 2006; Churazov
\etal 2005; Di~Matteo \etal 2005; Scannapieco \etal 2005; Bower \etal 2006;
Brighenti \& Mathews 2006; Croton \etal 2006; Hopkins \etal 2006; McNamara
\etal 2006); these studies maintain that AGN feedback plays a crucial role in
regulating the growth of early-type galaxies and their central supermassive
black holes.  Observations of giant elliptical galaxies in the local universe
have revealed significant \xray\ cavities filled with radio lobes that are
inferred to have been driven by nuclear AGN outbursts (e.g., Churazov \etal
2001, 2002; Jones \etal 2002, 2005; Forman \etal 2005; Nulsen \etal 2005a,
2005b; Allen \etal 2006); these observations suggest significant energy
transfer and recurrence timescales much shorter than the average cooling time.
One of the goals of this paper is to characterize the evolution of the hot
\xray-emitting gas in normal early-type galaxies and to place constraints on
the role of feedback by transient AGNs.

LMXBs provide another important contribution to the \xray\ emission from
early-type galaxies (both optically luminous and faint), and many
investigations have been dedicated to understanding their nature.  Observations
in the optical band, most notably from the {\it Hubble Space Telescope} (\hst),
have shown that many of the LMXBs in early-type galaxies are coincident with
globular clusters (e.g., Sarazin \etal 2003), and the number of LMXBs observed
in a given galaxy appears to be linearly correlated with the globular-cluster
specific frequency, $S_N$ (i.e., the number of globular clusters per unit
optical luminosity; e.g., White \etal 2002; Kim \& Fabbiano~2004; Irwin~2005;
Juett~2005).  White \etal (2002) examined several optically luminous,
LMXB-dominated early-type galaxies with stellar populations having ages ranging
from \hbox{$\approx$5--13~Gyr} (as measured via spectroscopy), and found that their
\hbox{X-ray--to--optical} luminosity ratios ($L_{\rm X}$[LMXB]$/L_{\rm opt}$) have no
dependence on galaxy age; $L_{\rm X}$(LMXB)$/L_{\rm opt}$ was observed to be
linearly proportional to $S_N$.  Again using optically luminous early-type
galaxies, Irwin (2005) found that details of the \hbox{$L_{\rm
X}$(LMXB)$/L_{\rm opt}$--$S_N$} relation imply that a significant number of
LMXBs were formed in the fields of the early-type galaxies and that the field
LMXB contribution to $L_{\rm X}$(LMXB)$/L_{\rm opt}$ is most significant, and
even dominates over the globular cluster LMXB emission, for galaxies with
smaller values of $S_N$.  In the optically luminous galaxies studied by White
\etal (2002) and Irwin (2005) $S_N$ is large, and LMXBs are mainly formed in
the high-density central regions of globular clusters through tidal
interactions between normal stars and either neutron stars or black holes.
In optically faint low-mass systems, however, $S_N$ is typically small (e.g.,
Ashman \& Zepf 1998), and it is therefore likely that LMXBs that formed within
the galactic fields of these systems dominate $L_{\rm X}$(LMXB)$/L_{\rm opt}$
(see, e.g., Fig.~3 of Irwin 2005); these LMXBs are thought to have formed via
the evolution of primordial binaries on timescales of \hbox{$\approx$1--10~Gyr}
following a star-formation event (e.g., Verbunt \& van~den~Heuvel~1995).
Furthermore, if the downsizing mass-assembly scenario discussed above is
correct, field LMXBs within optically faint galaxies could evolve significantly
over cosmic time, as LMXB populations emerge in the wake of relatively recent
bursts of star formation and eventually fade after $\approx$1~Gyr (e.g., Ghosh
\& White 2001).  Therefore, another goal of this paper is to measure the \xray\
properties of normal optically faint early-type galaxies over a significant
range of redshift to place constraints on the evolution of the LMXB activity
in these systems.

\subsection{Observations of the X-ray Emission from Distant Early-Type Galaxies}

At present, there have been few investigations of the redshift evolution of the
\xray\ properties of isolated early-type galaxies, which contrasts with the
case for late-type galaxies (e.g., Hornschemeier \etal 2002; Norman \etal 2004;
Laird \etal 2005; Kim \etal 2006; Lehmer \etal 2006).  Brand \etal (2005) used
\xray\ stacking analyses (see also, e.g., Brandt \etal 2001; Nandra \etal 2002;
Lehmer \etal 2005a) to investigate the evolution of optically luminous red
galaxies within the Bo\"{o}tes field of the NOAO Deep Wide-Field Survey (NDWFS)
over the redshift range \hbox{$z \sim$~0.3--0.9}.  The \chandra\ observations
used in these analyses were $\approx$5~ks in duration (e.g., Murray \etal 2005)
and covered an area of $\approx$1.4~deg$^2$.  Brand \etal (2005) found that the
average \xray\ luminosities of these galaxies mildly increase with redshift,
which is primarily due to a rise in AGN activity with redshift.  Due to its
relatively large sample size (3316 red galaxies), the Brand \etal (2005)
investigation provided useful statistical constraints on the \xray\ activity
from distant powerful AGNs; however, the \xray\ sensitivity of this study was
too low (i.e., the luminosity detection limit was \Lx~$\approx
10^{43.2}$~\xlum\ at $z \approx 0.7$) to distinguish effectively galactic
\xray\ emission (i.e., hot interstellar gas and LMXBs) from that of luminous
AGNs.  Deep \xray\ surveys, which utilize long \xray\ exposures with \chandra\
or \xmm\ (see, e.g., Brandt \& Hasinger 2005 for a review), are required to
investigate the \xray\ emission from distant ``normal'' (i.e., not
predominantly powered by luminous AGNs) early-type galaxies.  An ideally suited
field for such an investigation is the Extended \chandra\ Deep Field-South
(\hbox{E-CDF-S}).  The \ecdfs\ is a $\approx$0.3~deg$^2$, multiwavelength
survey field composed of the central $\approx$1~Ms \chandra\ Deep Field-South
(CDF-S; Giacconi \etal 2002; Alexander \etal 2003) and four flanking,
contiguous $\approx$250~ks \chandra\ observations (Lehmer \etal 2005b).  The
\hbox{0.5--2.0~keV} sensitivity limits (for detecting an individual unresolved
\xray\ point source) over the \ecdfs\ are $\approx$$5.2 \times 10^{-17}$~\flux\
in the most sensitive regions and $\approx$$3.0 \times 10^{-16}$~\flux\ over
the majority of the field; these levels correspond to \hbox{0.5--2.0~keV}
luminosity detection limits of $\approx$$10^{41.0}$ and
$\approx$$10^{41.5}$~\xlum, respectively, at $z = 0.7$.  Such sensitivity
limits are sufficient to detect moderately-powerful AGNs and normal, \xray\
luminous early-type galaxies (e.g., NGC~1600; Sivakoff \etal 2004) out to $z =
0.7$.

Recently, McIntosh \etal (2005; hereafter, M05) isolated a sample of 728
optically selected early-type galaxies within the \ecdfs\ from the GEMS (Galaxy
Evolution from Morphologies and SEDs; Rix \etal 2004) and \hbox{COMBO-17}
(Classifying Objects by Medium-Band Observations in 17 Filters; Wolf \etal
2004) surveys, which overlap with $\approx$0.23~deg$^2$ ($\approx$77\%) of the
\ecdfs.  The M05 sample spans a redshift range of \hbox{$z \approx$~0.2--1.0},
and due to its depth and relatively large comoving volume, it is ideal for
investigating the evolution of early-type galaxies over the last half of cosmic
history.  In this paper, we utilize the M05 sample and a supplementary sample
of 64 additional \hbox{$z \approx $~0.1--0.2} early-type galaxies located
within the \ecdfs\ (selected using the same techniques discussed in M05) to
characterize the redshift evolution of the \xray\ properties of early-type
galaxies.  Our aim is to measure the average \xray\ emission, via stacking
techniques (applied to \xray\ detected and \xray\ undetected sources), from
distant normal early-type galaxy populations.  We investigate separately the
cosmic-time evolution of (1) the hot interstellar gas content through optically
luminous galaxy samples and (2) the LMXB populations through optically faint
galaxy samples.  

The Galactic column density is \hbox{$N_{\rm H} \approx$~8.8 $\times$
10$^{19}$} cm$^{-2}$ for the \hbox{E-CDF-S} (Stark \etal 1992).  All of the
\hbox{X-ray} fluxes and luminosities quoted throughout this paper have been
corrected for Galactic absorption.  Unless stated otherwise, we make reference
to optical magnitudes using the Vega magnitude system.  $H_0$ = 70~\hbox{km
s$^{-1}$ Mpc$^{-1}$}, $\Omega_{\rm M}$ = 0.3, and $\Omega_{\Lambda}$ = 0.7 are
adopted throughout this paper (e.g., Spergel \etal 2003), which imply a
look-back time of 6.3~Gyr at $z=0.7$.

%
\section{Data Analysis}
%

\subsection{Sample Selection}

We started with the M05 sample of 728, \hbox{$z=$~0.2--1.0} early-type galaxies and a
supplementary sample of 64 \hbox{$z \approx$~0.1--0.2} galaxies within the
\ecdfs\ field (i.e., 792 total galaxies), which were selected following the
same methods outlined in M05 using the same data.  The total sample was
selected using (1) rest-frame $U-V$ red-sequence colors derived from
\hbox{COMBO-17} and (2) quantitative optical morphologies (via the S\'{e}rsic
indices; $n \ge 2.5$) as observed from \hst\ imaging through GEMS (see $\S$~2.3
of M05 for details).  The sample generated using these criteria is
representative of the early-type galaxy population as a whole and is highly
complete down to the selection limit of $R \approx 24$.  Photometric redshifts
for these early-type galaxies were provided by \hbox{COMBO-17} and are used
throughout this paper.  Using secure redshifts from various spectroscopic
surveys in the \ecdfs\ region (e.g., Le~Fevre \etal 2004; Szokoly \etal 2004;
Mignoli \etal 2005; Vanzella \etal 2005, 2006; Silverman \etal 2006), which
were available for $\approx$20\% of the M05 sample, we confirmed that the
photometric redshifts for our sample are highly accurate (i.e., $\approx$50\%
of the galaxies have $\delta z/[1+z_{\rm spec}] < 0.02$ and $\approx$75\% have
$\delta z/[1+z_{\rm spec}] < 0.03$).  The selection of early-type galaxies via
these quantitative methods isolates galaxies with visual morphologies ranging
from \hbox{E--Sa} galaxies, and a large majority of these are E/S0 galaxies (Bell
\etal 2004b).  We note that the \xray\ properties of E and S0 galaxies are
observed to be similar (e.g., Forman \etal 1985; Blanton \etal 2001; Xu \etal
2005).

Our primary goal was to investigate the redshift evolution of the \xray\
properties of normal early-type galaxies.  To achieve this better, we filtered
our original sample to include 539 galaxies within the redshift range \hbox{$z
\approx$~0.1--0.7}.  This range was chosen to optimize our use of the
\hbox{E-CDF-S} \hbox{X-ray} sensitivity to allow for the detection and
identification of low-luminosity AGNs over a broad range of cosmic look-back
times ($< 6.3$~Gyr).  As discussed in $\S$~1.3, the \chandra\ observations of the
\hbox{E-CDF-S} are sufficiently sensitive to detect AGNs with
\hbox{0.5--2.0~keV} luminosities \hbox{\Lx~$\simgt 10^{41.5}$~\xlum} over this
entire redshift range ($z \simlt 0.7$).  Furthermore, given that the
\hbox{X-ray} luminosities of AGN-hosting optically luminous ($L_B \simgt L_{B,
\rm crit}$) early-type galaxies in the local universe span \Lx~$\approx
10^{40.3-42.9}$~\xlum\ (mean \Lx~$\approx 10^{42.0}$~\xlum; O'Sullivan \etal
2001), this limit should adequately prevent powerful AGNs from dominating
stacked photon counts (assuming the AGN activity is not widespread).

\subsubsection{Removing AGN Contamination}

We utilized multiwavelength observations of the \ecdfs\ region to obtain a
census of the \xray-detected AGNs and normal galaxies within our sample.  We
matched galaxies in our sample to \xray\ sources in the catalogs presented in
Alexander \etal (2003)\footnote{See
http://www.astro.psu.edu/users/niel/hdf/hdf-chandra.html for the relevant
source catalogs and data products for the $\approx$1~Ms \cdfs.} for the
$\approx$1~Ms \cdfs\ observations and Lehmer \etal (2005b)\footnote{See
http://www.astro.psu.edu/users/niel/ecdfs/ecdfs-chandra.html for the relevant
source catalogs and data products for the $\approx$250~ks \ecdfs.} for the
$\approx$250~ks \ecdfs\ observations.  For a successful match, we required that
the optical and \xray\ centroids be displaced from each other by less than 1.5
$\times$ the radius of the \chandra\ positional error circles (\hbox{80--90\%}
confidence), which are given in each respective \xray\ catalog.  We note that
investigations of luminous off-nuclear \xray\ sources, which may lie outside
our matching radius, have shown that there is a dearth of such sources within
early-type galaxies (e.g., Irwin \etal 2004; Lehmer \etal 2006).  The \chandra\
source catalogs were generated using {\ttfamily wavdetect} (Freeman \etal 2002)
at false-positive probability thresholds of \hbox{$1 \times 10^{-7}$} and $1
\times 10^{-6}$ for the $\approx$1~Ms \cdfs\ and the $\approx$250~ks \ecdfs,
respectively.  However, as demonstrated in $\S$~3.4.2 of Alexander \etal (2003)
and $\S$~3.3.2 of Lehmer \etal (2005b), legitimate lower significance \xray\
sources, detected by running {\ttfamily wavdetect} at a false-positive
probability threshold of \hbox{$1 \times 10^{-5}$}, can be isolated by matching
with relatively bright optical sources.  Since the sky surface density of $z <
0.7$, $R < 24$ early-type galaxies is relatively low ($\approx$0.65 galaxies
arcmin$^{-2}$), we can effectively use this technique to isolate \xray\ sources
within our sample.  We estimate that when using {\ttfamily wavdetect} at a
false-positive probability threshold of \hbox{$1 \times 10^{-5}$}, we expect
$\approx$0.5 spurious sources.  In total, we detected 49 early-type galaxies in
at least one of the \hbox{0.5--2.0~keV}, \hbox{2--8~keV}, or
\hbox{0.5--8.0~keV} bandpasses. 

We used the following criteria to identify candidate AGNs within our sample:

\begin{enumerate}

\item {\it Hard X-ray Emission}: Our best discriminator of obscured ($N_{\rm H}
\simgt 10^{22}$~cm$^{-2}$) AGN activity is the presence of a hard \xray\
spectrum, which is characterized by a relatively shallow effective \xray\
spectral slope (i.e., $\Gamma_{\rm eff} \simlt 1.5$).  We classified \xray\
sources that were detected only in the \hbox{2--8~keV} bandpass as potential
AGN candidates.  For sources detected in both the \hbox{0.5--2.0~keV} and
\hbox{2--8~keV} bandpasses, we required that the hardness ratio measured using
these bandpasses, $\Phi_{\rm 2-8~keV}/\Phi_{\rm 0.5-2.0~keV}$ (where $\Phi$ is
the {\it observed} count-rate in each respective bandpass), be greater than 0.5
(corresponding to an effective photon index of $\Gamma_{\rm eff} \simlt 1.5$);
these sources would have spectral slopes shallower than those expected for a
pure LMXB population (e.g., Church \& Baluci{\'n}ska-Church 2001; Irwin \etal
2003).  We found that all sources with \hbox{2--8~keV} detections were
classified as AGN candidates.  Furthermore, there were a few sources that were
detected in only the \hbox{0.5--8.0~keV} bandpass.  We argue that since these
sources were not detected in the \hbox{0.5--2.0~keV} bandpass, our most
sensitive bandpass, then there must be a significant contribution from the
\hbox{2--8~keV} bandpass such that $\Phi_{\rm 2-8~keV}/\Phi_{\rm 0.5-2.0~keV} >
0.5$.  These sources were flagged as AGN candidates and removed from our
stacking analyses.  Using criterion~1, we identified a total of 29 AGN
candidates.

\item {\it X-ray--to--Optical Flux Ratio}:  Spectral hardness is a good
indicator of obscured AGN activity; however, powerful unobscured AGNs with
steep spectral slopes ($\Gamma \approx 2$) could still be missed if this were our
only means for identifying candidate AGNs.  Another useful discriminator of AGN
activity, which aids in the identification of luminous unobscured AGN
candidates, is the \hbox{0.5--8.0~keV} \hbox{\xray--to--optical} flux ratio, $f_{\rm
0.5-8.0~keV}/f_{\rm R}$ (e.g., Maccacaro \etal 1988; Hornschemeier \etal 2000;
Bauer \etal 2004).  We use the criterion $\log (f_{\rm 0.5-8.0~keV}/f_{\rm R})
> -1$ (see $\S$~4.1.1 of Bauer \etal 2004 for justification) to identify
unobscured AGN candidates in our sample; two candidates, which were not
identified by criterion~1 above, were found using this criterion.  We note that
16 out of the 29 AGNs that satisfied criterion~1 ($\approx$55\%) also
satisfied criterion~2.

\item {\it Radio--to--Optical Flux Ratio}:  We used radio maps (1.4~GHz;
limiting 1$\sigma$ depth of $\approx$14$\mu$Jy) from the Australia Telescope
Compact Array (ATCA; PI: A.~Koekemoer; Afonso \etal 2006), which cover the
entire \ecdfs\ region, to identify additional potential AGNs.  We matched the
positions of radio-detected sources to those of our early-type galaxies using a
2\farcs0 matching radius and found 18 matches.  Of the 18 radio-detected
sources in our sample, we find that three are detected in the \xray\ bandpass.
Of the three \xray-detected sources, only CXOECDFS \hbox{J033228.81--274355.6}
was not previously identified as an AGN candidate using criteria~1 and 2;
however, visual inspection of the radio source coincident with CXOECDFS
\hbox{J033228.81--274355.6} reveals that it has a clear FR~II radio morphology.
We find that the three AGN-like \xray-detected radio sources have relatively
large \hbox{radio--to--optical} flux ratios, $\log (\nu f_\nu [1.4\; {\rm
GHz}]/\nu f_\nu [5000$~\AA$])$, as expected for AGNs (e.g., Kinney \etal 1996).
Out of the remaining 15 radio-detected matches, 13 sources that were not
detected in the \xray\ bandpass were found to have uncharacteristically high
\hbox{radio--to--optical} flux ratios for early-type galaxies, $\log (\nu f_\nu
[1.4\; {\rm GHz}]/\nu f_\nu [5000$~\AA$])
> -4$.  These 13 sources were classified as potential AGNs and were excluded
from our stacking analyses described below.  We note that in addition to AGNs,
star-forming galaxies (e.g., edge-on dusty galaxies) that were misclassified as
early-type galaxies may also satisfy this criterion (see, e.g., Fig.~8 of
Barger \etal 2006); therefore, this crierion also helps to guard against the
inclusion of such sources.

\end{enumerate}

\noindent To summarize, we have detected 49 early-type galaxies in the \xray\
band and have identified 32 \xray-detected AGN candidates: 29 from criterion 1,
two additional sources from criterion 2, and one additional source from
criterion 3.  Finally, using criterion 3 we identified an additional 13 AGN
candidates that were not detected in the \xray\ band.  Therefore, we
identified a total of 45 AGN candidates from our sample of 539 early-type
galaxies.  The remaining 17 \xray-detected sources that we do not classify as
AGN candidates are considered to be normal early-type galaxies.  These normal
galaxies are included in our stacking analyses, and as we discuss in $\S$~2.2
below, including these galaxies does not significantly affect our results.  A
more detailed discussion of the \xray\ properties of our \xray-detected sources
is presented in $\S$~3.1 below.  We note that the three criteria discussed
above may not be completely sufficient to classify all \xray-detected sources
that are truly AGNs as AGN candidates (see, e.g., Peterson \etal 2006).  Such a
misclassification is possible for low-luminosity AGNs that are only detected in
the more sensitive \hbox{0.5--2.0~keV} bandpass, which have \hbox{2--8~keV}
emission too weak for an accurate classification.  Furthermore, we also expect
there to be AGNs that lie below the \xray\ detection threshold that are not
identified here.  However, in $\S$~3.2.2, we use the \hbox{2--8~keV} AGN
fraction as a function of \xray\ luminosity to argue quantitatively that we do
not expect misclassified AGNs (detected only in the \hbox{0.5--2.0~keV}
bandpass) and low-luminosity AGNs below the \xray\ detection limit to have a
serious impact on our results.  

\subsubsection{Normal Early-Type Galaxy Samples}

A large majority of the normal galaxies in our sample were not detected
individually in the deep \chandra\ observations over the \ecdfs.  We therefore
implemented stacking analyses (see $\S$~2.2 for a description) of galaxy
samples, which were constructed to cover two optical luminosity ranges in
multiple redshift bins.  We first divided the original sample of 539 early-type
galaxies by luminosity into optically luminous ($L_B \approx
10^{10-11}$~\lbsol) and optically faint ($L_B \approx 10^{9.3-10}$~\lbsol)
samples.  The two luminosity ranges were motivated by the observed physical
differences between local early-type galaxies within these two luminosity
ranges (see discussion in $\S$~1.1) and optical completeness limitations of the
M05 sample (i.e., the $R < 24$ completeness limit).  As mentioned in $\S$~1.1
and displayed in Figure~1, the optically luminous galaxies have soft \xray\
spectra dominated by hot interstellar gas, while the optically faint early-type
galaxies have \xray\ spectra powered mainly by LMXBs with a small contribution
from hot interstellar gas.  Galaxies from both luminosity classes have hard
\xray\ emission dominated by LMXBs.  

We divided the optically luminous sample into four redshift bins of equal
comoving volume (\hbox{$z \approx$~0.10--0.41}, \hbox{0.41--0.53},
\hbox{0.53--0.62}, and \hbox{0.62--0.70}), and due to optical completeness
limitations, we divided our optically faint sample into two redshift bins
(\hbox{$z \approx$~0.10--0.41} and \hbox{0.41--0.53}).  Although the
\hbox{E-CDF-S} subtends a relatively small solid angle, we note that each
redshift bin has a radial depth that exceeds 500~Mpc, which is a factor of
$\simgt$5 times larger than the extent of the largest structures in the
Universe (e.g., Springel \etal 2006).  We created histograms of the galaxy
luminosity-distance distributions on scales of $\approx$100~Mpc and did not
observe any dominating density ``spikes.''  This suggests that our samples are
not being significantly affected by cosmic variance.  Furthermore, provided the
physical properties of field early-type galaxies are not strongly affected by
their environments in a systematic way, we suggest that cosmic variance should
not affect our results even if present.  In Figure~2a, we show rest-frame
$B$-band luminosity, $L_B$ (in solar units; \lbsol~$= 5.2 \times
10^{32}$~\xlum), versus redshift for the 539 galaxies ({\it dots} and {\it
crosses\/}) that make up our ``general sample''; we indicate the divisions of
the sample by optical luminosity ({\it shaded regions\/}) and redshift ({\it
vertical dotted lines\/}).  Sources highlighted by large circles (both {\it
open} and {\it filled\/}) are those that are individually detected in
\hbox{X-rays}; the open circles indicate \hbox{X-ray}-detected AGN candidates,
and the filled circles indicate the remaining detected sources, which are
classified as normal early-type galaxies.  

In addition to the general sample, we have created ``faded samples'' of
early-type galaxies.  The faded samples were constructed using the general
sample discussed above; however, we have corrected the rest-frame $B$-band
luminosities for passive evolution.  These corrections take into account the
fact that the optical power output of early-type galaxies has been fading with
cosmic time due to stellar evolution (see $\S$~1).  For each galaxy, we have
computed an evolved, $z=0$ $B$-band luminosity ($L_{B,0}$) following the
best-fit optical-size-dependent solutions for optical fading presented in
Tables~1 and 2 of M05.  M05 present eight best-fit relations for $\Delta M_V
(z)$ (i.e., the difference between the absolute $V$-band magnitude at redshifts
$z$ and $z=0$) versus $z$, which differ in their evolutionary scenarios (based
upon formation scenarios and the {\ttfamily PEGASE} models of Fioc \&
Rocca-Volmerange~1997) and whether or not the fit is constrained to $\Delta M_V
(z=0) = 0$.  Throughout this paper we quote results based upon the best-fit
solution that assumes $\Delta M_V (z=0) = 0$ and a single-burst evolutionary
model with a formation redshift of $z_{\rm form} = 3$ and a metallicity of
\hbox{[Fe/H] = $-$0.2} (see $\S$~3.2 of M05 for details).  We converted the $V$-band
relation to the $B$-band following $\Delta M_B \approx 1.1 \times
\Delta M_V$.  Additional \xray\ analyses, using the seven alternative
evolutionary scenarios presented in M05, were also performed, but no material
differences were observed in our results.  Figure~2b shows $L_{B,0}$ versus $z$
for our faded sample; symbols, lines, and shaded regions have the same meaning
as in Figure~2a.  In generating samples for stacking, we made the same sample
divisions (of both luminosity and redshift) as we did for the general sample,
except we used $L_{B,0}$ to discriminate between optically luminous and faint
galaxies.  This approach allows us to place constraints on the evolution of the
\xray\ activity from distant galaxies selected from the optical band as they
would appear today.  Furthermore, based on local estimates of stellar
\hbox{mass--to--light} ratios, we can use $L_{B,0}$ to estimate the stellar masses of
these systems.  Using a Chabrier initial mass function (Chabrier 2003), we
estimate that the luminosity ranges $L_{B,0} = 10^{9.3-10}$ and
$10^{10-11}$~\lbsol\ correspond roughly to stellar-mass ranges of $\approx
10^{9.9-10.6}$ and $\approx 10^{10.6-11.6}$~\msol, respectively.  In the
presentation below, we focus our attention on results drawn from the faded
samples since these are expected to be the most physically meaningful.

\subsection{X-ray Stacking Technique}

In order to address the fact that the \xray\ emission from early-type galaxies
is dominated by different physical processes in different energy bands, we
performed stacking analyses in three bandpasses: \hbox{0.5--1.0}~keV
(SB1),\footnote{The \hbox{0.5--1.0~keV} bandpass was originally defined as SB1
in $\S$~3.1 of Alexander \etal (2003).} \hbox{0.5--2.0~keV} (soft band; SB),
and \hbox{2--8~keV} (hard band; HB).  The shaded bars in the panels of Figure~1
show the spectral coverage of these bandpasses at the median redshifts of our
optically luminous ($z_{\rm median} \approx 0.55$) and faint ($z_{\rm median}
\approx 0.39$) samples for the \xray\ spectral energy distributions (SEDs) of
NGC~1600 ({\it top panel\/}) and NGC~4697 ({\it bottom panel\/}), respectively.
For optically luminous galaxies, the SB1 and SB will effectively sample \xray\
emission dominated by hot interstellar gas, which produces $\approx$80\% and
$\approx$70\% of the total emission in each respective bandpass.  In contrast,
90\% of the HB flux originates from LMXB emission.  For optically faint
galaxies, SB and HB will effectively sample the \xray\ emission from LMXBs;
however, the \xray\ emission observed in SB1 has roughly equal
contributions from hot gas and LMXB emission.  In our analyses, we used data
products from Alexander \etal (2003) for the $\approx$1~Ms \cdfs\ and Lehmer
\etal (2005b) for the $\approx$250~ks \ecdfs\ (see footnotes~1 and 2).  The
data products are publicly available for all energy bands except for SB1 in the
$\approx$250~ks \ecdfs\ regions, which were generated specifically for the
analyses here using the same methods described in Lehmer \etal (2005b).  The
methodology of our stacking procedure was similar to that outlined in Lehmer
\etal (2005a); for completeness, we describe this procedure below.

We maximized our stacked signal by optimizing our choices of circular
stacking aperture radius (from which we extract photon counts for both sources
and their backgrounds) and inclusion radius (i.e., the maximum off-axis angle
within which we include sources for stacking).  This process is needed
because the \chandra\ point spread function (PSF) increases in size with
off-axis angle, which degrades the sensitivity for those sources that are far
off-axis.  For the optimization process, we stacked all early-type
galaxies in our sample that were not detected individually in the \xray\ band.
In order to obtain a clean \xray\ signal, we excluded galaxies that were located
$\simlt$10\arcsec\ from unrelated sources in the \xray\ source
catalogs and within the extent of extended \xray\ sources, which are likely
associated with galaxy groups or poor clusters (see $\S$~3.4 of Giacconi \etal
2002 and $\S$~6 of Lehmer \etal 2005b).  Sources that lie within both the
$\approx$1~Ms \cdfs\ and $\approx$250~ks \ecdfs\ were stacked using both
observations, as long as the off-axis angle for each observation was within our
chosen inclusion radius.  

Our optimization procedure was a two-step iterative process that was performed
using the SB, the bandpass in which our signal was maximized.  In the first
step of the optimization process ({\it step one\/}) we held the inclusion
radius ($R_{\rm incl}$) fixed and stacked sources using a variety of circular
stacking apertures of constant radii (i.e., we did not vary the aperture radius
as a function of off-axis angle).  We used 15 different circular stacking
apertures with radii in the range of \hbox{0\farcs5--3\farcs0} to obtain a
relation for how the signal-to-noise ratio (S/N) varied as a function of
aperture size.  For a given aperture, we stacked the photon counts and
effective exposure times from each galaxy position and summed them to obtain
total source-plus-background counts, $S$, and a total exposure time, $T$,
respectively.  We estimated total background counts, $B$, using Monte Carlo
simulations and the background maps described in $\S$~4.2 of Alexander \etal
(2003) for the $\approx$1~Ms \cdfs\ and $\S$~4 of Lehmer \etal (2005b) for the
$\approx$250~ks \ecdfs.  We shifted our aperture from each galaxy position
randomly within a circular region of radiud $\approx$25\arcsec, extracted
background counts from each relevant background map, and summed the counts to
obtain an estimate of the total background counts.  This procedure was repeated
10,000 times to obtain an estimate of the local background and its dispersion.
Our best estimate of the total background counts, $B$, was approximated by
taking the mean background calculated from the 10,000 Monte Carlo trials.  For
each of the 15 circular apertures, we computed the relevant signal-to-noise
ratio (S/N) using the equation, S/N $\equiv (S-B)/\sqrt{B}$.  After performing
the stacking procedure for all 15 different apertures, we identified the
aperture radius that produced the maximal S/N.  In the second step of the
optimization process ({\it step two\/}), we held the optimized aperture
determined from step one fixed, but this time we stacked sources using 15
different inclusion radii ranging from \hbox{1\farcm5--11\farcm0} to obtain S/N
as a function of inclusion radius.  After performing the stacking procedure for
all 15 different inclusion radii, we identified the optimal inclusion radius.
For a clean stacking signal, we expect that S/N will be proportional to the
inclusion radius (i.e., S/N~$\propto [{\rm Number \; of \; sources}]^{1/2}
\propto [\pi R_{\rm incl}^2]^{1/2} \propto R_{\rm incl}$), and therefore we
chose the optimal inclusion radius based on where the \hbox{S/N--$R_{\rm
incl}$} relation appears to deviate significantly from linear.  This method
helps to guard against the inclusion of contaminating AGNs in the
low-sensitivity regions at large off-axis angles.  Using the optimal inclusion
radius determined with this method, we repeated step one.  This two-step
process was run iteratively until a converged solution was obtained.  We found
that a stacking aperture of $\approx$1\farcs5 and an inclusion radius of
$\approx$7\farcm0 produced the optimal signal; we note that this choice of
optimized $R_{\rm incl}$ is somewhat smaller than that determined in Lehmer
\etal (2005a) due to differing \xray\ exposures and the more conservative
approach that we have adopted for choosing the optimal $R_{\rm incl}$.
Hereafter, we utilize these values for stacking aperture and inclusion radius
in our analyses.  In Figure~3, we show S/N as a function of inclusion radius
(Fig.~3a) and aperture size (Fig.~3b), which were obtained by holding the
aperture radius constant at 1\farcs5 and the inclusion radius constant at
7\farcm0, respectively.

After excluding sources that were (1) classified as AGN candidates (via the
three criteria outlined in $\S$~2.1.1), (2) located at off-axis angles greater
than 7\farcm0, (3) within regions of extended \xray\ emission, and (4) within
10\arcsec\ of an unrelated \xray-detected sources, we were left with general
and faded samples of 276 and 229 galaxies, respectively; in both samples, 13
galaxies were \xray-detected sources. Using these samples, we stacked the \xray\
properties following the procedure described above.  The sizes of the general
and faded samples differ most notably because of the fading of high-redshift
optically luminous galaxies out of the selected luminosity range.   We stacked
samples both with and without \xray-detected normal galaxies included, and no
material differences were observed in our results; therefore, \xray-detected
sources that were classified as normal galaxies were included in our stacking
analyses.  Figure~4 shows the spatial distribution of the 229 stacked sources
in our faded sample over the \ecdfs\ region.  Sources stacked using the
$\approx$1~Ms and $\approx$250~ks observations are shown as circles and
diamonds, respectively; the sources indicated with filled circles were stacked
using both observations.  For each stacked sample, we required the S/N be
greater than or equal to 3 (i.e., $\simgt$99.9\% confidence) for a detection.
For stacked samples without significant detections, 3$\sigma$ upper-limits were
placed on the source counts.  

Using net counts (i.e., $S-B$) from our stacked samples and adopted \xray\
SEDs (see below), we calculated absorption-corrected fluxes and rest-frame
luminosities.  Due to the fact that our 1\farcs5 radius stacking aperture
encircles only a fraction of the PSF\footnote{For SB and SB1, the
encircled-energy fraction of a 1\farcs5 radius circular aperture varies from
$\approx$100\% at off-axis angle $\theta \simlt 3\arcmin$ to $\approx$45\% at
$\theta \approx 7\arcmin$.  For the HB, this fraction varies from $\approx$80\%
at $\theta \simlt 3\arcmin$ to $\approx$40\% at $\theta \approx 7\arcmin$.} for
sources at relatively large off-axis angle, we calculated aperture corrections
$\xi_i$ for each stacked source $i$.  

Since we are calculating average \xray\
counts from the summed emission of many sources of differing backgrounds and
exposure times, we used a single, representative exposure-time-weighted
aperture correction, $\xi$.  This factor, which was determined for each stacked
sample, was calculated following:

\begin{equation}
\xi \equiv \frac{\sum_i \xi_i \times T_i}{\sum_i T_i},
\end{equation}

\noindent where $T_i$ is the effective exposure time for each stacked source.
The average aperture corrections ($\xi$) for our samples were $\approx$1.5 and
$\approx$1.8 for the \hbox{0.5--2.0~keV} and \hbox{2--8~keV} bands,
respectively.  We note that at the mean redshifts of our samples, the
$\approx$1\farcs5 radius aperture corresponds to projected physical radii in
the range of \hbox{5.8--10.5~kpc}.  For early-type galaxies in the local
universe, extended \xray\ emission originating outside of these radii varies
considerably among galaxies and generally represents only a small fraction
(\hbox{$\approx$1--10\%}) of the total flux (e.g., Fukazawa \etal 2006).
Therefore, we do not make additional aperture corrections to account for
extended \xray\ emission.  We estimated mean observed count rates ($\Phi$)
using the following equation:

\begin{equation}
\Phi = \xi (S-B)/T,
\end{equation}

\noindent where $S$, $B$, and $T$ are defined above.  To convert count rate to
flux, we used the Galactic column density given in $\S$~1.3 and the best-fit
\xray\ spectral energy distribution (SED) for NGC~1600 for the optically
luminous galaxies (see Fig.~1a) and a power-law SED with $\Gamma = 1.8$ for the
optically faint galaxies (see $\S$~1.1 for justification).  We note that for a
range of reasonable SED choices, we find systematic fractional uncertainties in
the count-rate to flux conversion of $\simlt$10\% for SB and HB and
$\approx$50\% for SB1.  Mean rest-frame \xray\ luminosities $L_{\rm X,R}$ were
calculated following:

\begin{equation}
\langle L_{\rm X,R} \rangle \approx 4 \pi \langle d^2_L \rangle \langle f_{\rm X,O} \rangle  K,
\end{equation}

\noindent where $d_L$ is the luminosity distance, $\langle f_{\rm X,O} \rangle$ is the mean
observed-frame \xray\ flux, and $K$ is a unitless conversion factor, which
relates the observed-frame \xray\ flux to the rest-frame luminosity using our
adopted SEDs.  The fractional errors on $\langle L_{\rm X,R} \rangle$ due to uncertainties in
$\langle d_L^2 \rangle$ range from $\approx$10\% to $\approx$1\% for our optically luminous
\hbox{$z \approx$~0.10--0.41} and \hbox{$z \approx$~0.62--0.70} samples, respectively.  Due
to the relatively broad energy ranges our bandpasses encompass and the
relatively small spectral shifts observed over the redshifts of sources in our
samples, we used observed-frame fluxes to compute rest-frame luminosities.  For
our adopted SEDs, we found that values of $K$ varied between 0.8 and 1.1,
depending on the bandpass and redshift of the source.  For all bandpasses, SED
choice contributes small systematic uncertainties in $K$, which range from
\hbox{$\approx$10--20\%} depending on redshift.  We note that the mean \xray\
luminosity is expected to be closely representative of a typical galaxy within
the confined optical luminosity ranges used here.  Moreover, using the
O'Sullivan \etal (2001) sample of local early-type galaxies, we find that
$L_{\rm X}^{\rm mean}/L_{\rm X}^{\rm median} \approx 1.7$ and 1.2 for optically
luminous and faint samples respectively.  Since we are only able to calculate
mean quantities via the \xray\ stacking used here, we utilize mean quantities
throughout.  Results from our stacking analyses are presented below in
$\S$~3.2.

\section{Results}

\subsection{Individually Detected \xray\ Sources}

In Table~1, we present the properties of the \xray-detected early-type galaxies.
\xray\ source properties were determined following
the methods outlined in Alexander \etal (2003) and Lehmer \etal (2005b) for the
\cdfs\ and \ecdfs, respectively.  Using the matching criterion discussed in
$\S$~2.1.1, we matched 39 of our early-type galaxies to sources that were
included in the main \chandra\ catalogs of either Alexander \etal (2003) or
Lehmer \etal (2005b), and an additional ten matches were identified using
\xray\ sources detected using {\ttfamily wavdetect} at a false-positive
probability threshold of $1 \times 10^{-5}$ (i.e., 49 total detected sources).
Figure~5 shows the $R$-band magnitude (from COMBO-17) versus
\hbox{0.5--8.0~keV} flux for all 49 \xray\ detected early-type galaxies in our
sample; normal galaxies and AGN candidates are indicated as filled and open
circles, respectively.  As discussed in $\S$~2.1.1, we classified 32
($\approx$65\%) \xray-detected sources as AGN candidates.  We note that the
majority of the AGN candidates have $\log f_{\rm 0.5-8.0~keV}/f_{\rm R} > -1$;
however, a few of these candidates have $\log f_{\rm 0.5-8.0~keV}/f_{\rm R} <
-2$, including the FR~II source CXOECDFS \hbox{J033228.81--274355.6}.  AGNs
with $\log f_{\rm 0.5-8.0~keV}/f_{\rm R} < -2$ are either significantly
obscured or relatively \xray-weak AGNs.

As discussed in $\S$~1.2, transient AGN feedback may play an important role
in heating the hot gas in normal early-type galaxies.  In order to understand
and constrain the influence of transient AGN activity within early-type
galaxies, we computed the \xray-luminosity-dependent cumulative AGN fraction,
$f_C$ (i.e., the fraction of early-type galaxies harboring an AGN with a
\hbox{2--8~keV} luminosity of $L_{\rm 2-8~keV}$ or greater).  We made use of
the \hbox{2--8~keV} bandpass because of its ability to probe relatively
unattenuated \xray\ emission in a regime of the \xray\ spectrum where we expect
there to be minimal emission from normal galaxies (see also $\S$~2.1.1 for more
details).  Figure~2 illustrates that the majority of the AGNs in our samples
({\it open circles\/}) originate within optically luminous ($L_B \simgt
10^{10}$~\lbsol) early-type galaxies, and therefore when computing $f_C$, we
used only optically luminous galaxies; we note that the number of AGNs within
our optically faint samples is too low to obtain a respectable constraint on
$f_C$.  In order to quantify the redshift evolution of $f_C$, we split our
optically luminous samples into two redshift intervals of roughly equal size
($z \approx$~\hbox{0.10--0.55} [$z_{\rm median}=0.42$] and $z
\approx$~\hbox{0.55--0.70} [$z_{\rm median}=0.65$]).  Only two redshift
intervals were chosen due to statistical limitations on the number of detected
AGNs.  We computed $f_C$ by taking the number of candidate AGNs with a
\hbox{2--8~keV} luminosity of $L_{\rm 2-8~keV}$ or greater and dividing it by
the number of early-type galaxies in which we could have detected an AGN with
luminosity $L_{\rm 2-8~keV}$.  The latter number was computed by considering
the redshift of each galaxy and its corresponding sensitivity limit, as
obtained from spatially dependent sensitivity maps (see $\S$~4.2 of Alexander
\etal 2003 and $\S$~4 of Lehmer \etal 2005b); these sensitivity maps were
calibrated empirically using sources detected by {\ttfamily wavdetect} at a
false-positive probability threshold of $1 \times 10^{-5}$.  Figure~6 shows
$f_C$ as a function of $L_{\rm 2-8~keV}$ for the two redshift bins considered
here.  We find suggestive evidence for evolution in $f_C$ between $z
\approx$~0.42 and $z \approx$~0.65, which is consistent with the global trend
observed for luminous AGNs in general (e.g., Brandt \& Hasinger 2005).  At
$L_{\rm 2-8~keV} \simgt 10^{42.2}$~\xlum, where $f_C$ is most tightly
constrained, we find that $f_C(z=0.65)= [2.1_{-2.0}^{+10.5}] \times
f_C(z=0.42)$.  We note that although this value is poorly constrained, it is
consistent with the $(1+z)^3$ evolution observed for \xray\ luminosity
functions of \xray-selected AGNs (e.g., Ueda \etal 2003; Barger \etal 2005;
Hasinger \etal 2005), and is in agreement with the stacked constraints on
optically-selected early-type galaxies set by Brand \etal (2005).  We return to
the discussion of \xray-detected AGNs in $\S$~3.2.2 when discussing the
undetected AGN contribution to our stacked samples and in $\S$~4.1 when
discussing the transient AGN contribution to heating the hot interstellar gas
in optically luminous early-type galaxies.

\subsection{X-ray Stacking Results}

\subsubsection{Stacked Properties}

In Tables~2 and 3, we summarize the average properties of our stacked samples
of normal early-type galaxies.  For illustrative purposes, we created Figure~7,
which shows \hbox{0.5--2.0~keV} (SB) adaptively-smoothed stacked images of our
faded samples.  We detect the average \xray\ emission from all of our samples
in SB, several in SB1, and only two in HB.  The two samples from which we
detect HB emission (the $z \approx 0.65$ optically luminous general sample and
the $z \approx 0.58$ optically luminous faded sample) have \hbox{HB--to--SB1}
and \hbox{HB--to--SB} count-rate ratios (i.e., $\Phi_{\rm 2-8~keV}/\Phi_{\rm
0.5-1.0~keV}$ and $\Phi_{\rm 2-8~keV}/\Phi_{\rm 0.5-2.0~keV}$) that are broadly
consistent with our adopted \xray\ SED.  Furthermore, all of our stacked
samples have mean \hbox{\xray--to--optical} flux ratios ($f_{\rm
0.5-8.0~keV}/f_{\rm R}$) that are consistent with those expected for normal
galaxies (see col.[5] of Table~3 and Fig.~5).  In Figure~5, we have plotted
mean quantities from our samples as large filled squares and triangles, which
represent our optically luminous and faint faded samples, respectively; these
symbols have been shaded with varying grayscale levels to indicate redshift,
such that darker shading represents higher redshift samples.

For the purpose of comparing our results with local early-type galaxies, we
utilize the O'Sullivan \etal (2001; hereafter OS01) sample.  The OS01 sample
was selected from the Lyon-Meudon Extragalactic Data Archive (LEDA) using
morphological type ($T < -1.5$; \hbox{E--S0} Hubble types), distance ($V \le
9000$~km~s$^{-1}$), and apparent magnitude ($B_T \le 13.5$).  The LEDA catalog
is known to be $\approx$90\% complete down to $B_T = 14$.  \xray\ observations
of these galaxies were available mainly from the \rosat\ PSPC with a
significant minority of the data originating from the \einstein\ IPC.  We also
utilized \chandra\ data from David \etal (2006) for six of the OS01 galaxies
having only \xray\ upper limits.  Figure~8 shows the \hbox{0.5--2.0~keV}
luminosity versus $B$-band luminosity for galaxies included in the OS01 sample
with $D < 70$~kpc; \xray\ luminosities have been normalized to the
\hbox{0.5--2.0~keV} bandpass using the \xray\ SED adopted in OS01 (i.e., a
{\ttfamily MEKAL} plasma SED with solar metallicity and a plasma temperature of
1~keV).  We have denoted field, central-cluster, brightest-group, and
AGN-hosting early-type galaxies as identified from the OS01 sample; several
well-studied examples have been highlighted (M32, M87, NGC~1399, NGC~1600,
NGC~4697, and NGC~5102) for reference.  

As noted in $\S$~1.1, the \xray\ and $B$-band luminosities of local early-type
galaxies are observed to be correlated, and these correlations follow a power
law:

\begin{equation}
\log L_{\rm X} = \alpha \log L_B + \beta,
\end{equation}

\noindent where $\alpha$ and $\beta$ are fitting constants.  Using the OS01
sample, we performed linear-regression analyses for galaxies with $L_B \simgt
10^{10}$~\lbsol\ and \hbox{$L_B \simlt 10^{10}$}~\lbsol\ separately to
determine $\alpha$ and $\beta$ for each luminosity regime.  When doing these
calculations, we excluded (1) sources at $D > 70$~kpc, (2) sources with \xray\
emission that may be significantly influenced by \xray-emitting gas associated
with galaxy clusters or groups such as central-cluster and brightest-group
galaxies, (3) AGN-hosting galaxies, (4) NGC~5102, due to its anomalously low
\xray\ luminosity and evidently recent star-formation activity (e.g., OS01;
Kraft \etal 2005; see Fig.~8), and (5) NGC~4782, which has an anomalously large
$L_B$ that drives the correlation (i.e., $L_B = 10^{11.4}$~\lbsol; see, e.g.,
OS01).  We utilized Kendall's tau (Kendall 1938) to measure the correlation
strengths and Buckley-James regression (Buckley \& James 1979; Isobe \etal
1986) to calculate the best-fit correlation parameters (i.e., $\alpha$ and
$\beta$) for each luminosity regime.  These tools were available through the
Astronomy SURVival Analysis software package ({\ttfamily ASURV} Rev.~1.2; Isobe
\& Feigelson 1990; LaValley \etal 1992).  We found correlation significances of
$\approx$6.8$\sigma$ and $\approx$2.8$\sigma$ for the $L_B \simgt
10^{10}$~\lbsol\ and \hbox{$L_B \simlt 10^{10}$}~\lbsol\ galaxy samples,
respectively.  We found that for the optically luminous sample \hbox{$\alpha =
2.61 \pm 0.61$} and \hbox{$\beta = 12.77$} ({\it solid line\/} in Fig.~8), and
for the optically faint sample \hbox{$\alpha = 1.05 \pm 0.27$} and \hbox{$\beta
= 29.36$} ({\it dashed line\/} in Fig.~8).  We note that the \hbox{\Lx--$L_B$}
relation for optically faint galaxies is poorly constrained with $\approx$75\%
of the galaxies having only \xray\ upper limits; however, since these galaxies
are thought to be dominated by LMXBs, investigations of the LMXB luminosity per
unit $B$-band luminosity can offer a consistency check for this relation.
Using \chandra\ observations of 14 E/S0 galaxies, Kim \& Fabbiano (2004) found
that for LMXBs with \Lx~$> 10^{37}$~\xlum, $\beta = 29.5 \pm 0.25$ using a
fixed slope of $\alpha = 1$.  This value is the most tightly constrained
\hbox{\Lx(LMXB)--$L_B$} relation available at present and is consistent with other
investigations of the discrete-source contribution (e.g., Sarazin 1997; OS01
and references therein; Gilfanov~2004) and our calculated \hbox{\Lx--$L_B$}
relation for local optically faint early-type galaxies.  For reference, the
expected discrete-source contribution from LMXBs is presented in Figure~8 as a
shaded region, which represents the dispersion of the relation.  We note that
the \hbox{\Lx--$L_B$} relation for local optically faint early-type galaxies
({\it dashed line\/}) is consistent with that expected for \xray\ emission
originating strictly from LMXBs.

In Figure~8, we have plotted mean quantities from our samples using the same
symbols and symbol-shading schemes that were adopted in Figure~5.  The plotted
error bars for our stacked samples were computed by propagating (1) Poisson
errors on the source counts (Gehrels~1986), (2) 1~$\sigma$ errors on $\langle
d_L^2 \rangle$, and (3) systematic errors on the SED-dependent parameters
(i.e., count-rate to flux conversion and $K$); these errors were propagated
following the ``numerical method'' described in $\S$1.7.3 of Lyons (1991).  In
an initial evaluation of these results, we find that the average \xray\
properties of our optically luminous samples appear to follow the local
relation at all redshifts.  In contrast, our optically faint samples deviate
from the local relation significantly (1.5$\sigma$ at $z \approx 0.46$) over
the redshift range \hbox{$z \approx$~0.1--0.5}, suggesting there may be some
evolution in the LMXB populations within these galaxies. 

Figures~9a and 9b show the mean \hbox{0.5--2.0~keV} luminosities (\Lx) of our
stacked general and faded samples, respectively; optically luminous and faint
samples are indicated as dark filled squares and triangles, respectively.
\xray-detected sources are shown as circles, which represent both normal
galaxies ({\it filled circles\/}) and AGN candidates ({\it open circles\/}).
We have included the corresponding mean \xray\ luminosities for local
early-type galaxies from the OS01 optically luminous and faint samples in
matched optical-luminosity ranges; as before, these samples were filtered to
exclude galaxies classified as AGNs, central-cluster galaxies, brightest-group
galaxies, NGC~5102, and NGC~4782.  Mean optical and \xray\ luminosities for the
OS01 samples were computed using the Kaplan-Meier estimator (e.g., Feigelson \&
Nelson 1986) within {\ttfamily ASURV}, which appropriately handles censored
data.  When calculating these mean luminosities, we filtered the OS01 sample to
include only sources with distances $< 40$~Mpc and $< 20$~Mpc for the optically
luminous and faint samples, respectively; these distances represent approximate
completeness limits for the optical-luminosity ranges used here.  We calculated
mean \xray\ luminosities of \hbox{$\log \langle L_{\rm X} \rangle \approx 40.2
\pm 0.2$} and $\approx$$39.2 \pm 0.2$ for the optically luminous and faint
samples, respectively; error bars here represent 1$\sigma$ errors on the means
(computed with {\ttfamily ASURV}).  These calculations were made using 102
optically luminous galaxies (60 upper limits) and 48 optically faint galaxies
(36 upper limits).  We attempted to improve the \xray-detection fractions of
these samples by using subsamples of galaxies that were created with distance
limits smaller than the completeness limits quoted above; however, we found no
improvement in these fractions.

Figures~10a and 10b show the \hbox{\xray--to--$B$-band} mean luminosity ratio ($L_{\rm
X}/L_{B}$) for our general and faded samples, respectively (symbols have the
same meaning as they did in Fig.~9); the expected local discrete-source
contribution and its dispersion are shown as a horizontal dotted line and the
surrounding shaded region, respectively.  As observed in Figure~8, we find
little evidence for evolution in our optically luminous samples (both general
and faded).  Using the available data, we find that $(L_{\rm X}/L_{B})_{z=0.7}
= [1.0 \pm 0.5]$ and $[1.7 \pm 0.8] \times (L_{\rm X}/L_{B})_{z=0}$ for our
optically luminous general and faded samples, respectively.  In order to
constrain further the allowed redshift evolution of $L_{\rm X}/L_{B}$, we
utilized the $\chi^2$ statistic and a simple evolutionary model, $(L_{\rm
X}/L_{B})_{z} = (1+z)^\eta (L_{\rm X}/L_{B})_{z=0}$.  For this single-parameter
model, we constrained $\eta$ using 90\% confidence errors ($\Delta \chi^2
=2.7$).  We found best-fit parameters of $\eta = -0.4^{+0.6}_{-0.7}$ ($\chi^2 =
3.4$ for 4 degrees of freedom) and $\eta = 0.4^{+0.6}_{-0.7}$ ($\chi^2 = 11.3$
for 4 degrees of freedom) for our optically luminous general and faded samples,
respectively.  For the optically faint early-type samples, we observe
suggestive redshift evolution in $L_{\rm X}/L_{B}$, and by $z \approx 0.5$ it
has increased above the local relation by a factor of $\approx$$5.3 \pm 4.1$
and $\approx$$5.6 \pm 4.1$ for our general and faded samples, respectively.
Using the $\chi^2$ statistic and the same simple model for redshift evolution
presented above, we find best-fit values of $\eta = 4.4 \pm 1.7$ ($\chi^2 = 1.1
\times 10^{-3}$ for 2 degrees of freedom) and $\eta = 5.4 \pm 1.5$ ($\chi^2 =
1.0$ for 2 degrees of freedom) for our optically faint general and faded
samples, respectively.  We note that the evolution observed for optically faint
early-type galaxies is largely driven by the value of $L_{\rm X}/L_{B}$ at
$z=0$.  Due to the fact that the $z=0$ value for $L_{\rm X}/L_{B}$ is based on
48 sources, with 36 (75\%) having only \xray\ upper limits, we cannot rule out
the possibility that $L_{\rm X}/L_{B}$ at $z=0$ is significantly affected by
systematic errors.  Moreover, the total \xray\ emission from optically faint
galaxies is expected to vary significantly between galaxies due to low numbers
of LMXBs and variable amounts of hot gas, and therefore large fractional errors
are expected for $L_{\rm X}/L_{B}$.  We therefore consider this result to be
only marginal at present.  

As noted above, the local relation between \Lx\ and $L_B$ is nonlinear for
optically luminous early-type galaxies.  In order to investigate whether such
nonlinearities have an effect on our overall results, we created Figure~11,
which illustrates residuals to the local best-fit relations, $\log L_{\rm X} =
\alpha \log L_B + \beta$, for optically luminous ($\alpha=2.61$; $\beta=12.77$)
and faint ($\alpha=1.05$; $\beta=29.36$) samples.  Figure~11 shows that the
nonlinearities observed in the local \hbox{\Lx--$L_B$} relation do not affect our
conclusions drawn from using $L_{\rm X}/L_{B}$ as a proxy for evolution.
Furthermore, due to the approximate equality of the mean values of $L_B$ for
all samples of a given luminosity class, quantitative analyses that account for
nonlinearities in the \hbox{$L_{\rm X}$--$L_B$} relations yield roughly identical
results to those quoted using simply \hbox{$L_{\rm X}/L_{B}$}.

\subsubsection{Assessing Remaining AGN Contamination}

In this section we assess whether the stacked properties presented above suffer
from contamination by AGNs with \hbox{2--8~keV} luminosities below our
detection threshold.  In $\S$~3.1 (see also Fig.~6), we presented the
cumulative AGN fraction, $f_C$, the fraction of early-type galaxies harboring
an AGN with a \hbox{2--8~keV} luminosity of $L_{\rm 2-8~keV}$ or greater.  To
first order, we can use the functional form of $f_C$ to generate a census of
the AGN population that we expect to be missing due to sensitivity limitations.
As noted in $\S$~3.1, there is evidence that $f_C$ evolves with redshift.  We
modelled this redshift evolution of $f_C$ using the functional form, $f_C(z) =
(1+z)^3 f_C(z=0)$ (see $\S$~3.1 for justification).  We also assumed that the
dependence of $f_C$ upon $L_{\rm 2-8~keV}$ was similar at all redshifts over
the range $z \approx$~\hbox{0.1--0.7}.  Using our faded sample of optically
luminous early-type galaxies we computed $\langle f_C \rangle _z$, the average
cumulative AGN fraction over the redshift range \hbox{$z \approx$~0.1--0.7}
($z_{\rm medan}=0.55$), following the procedure outlined in $\S$~3.1; this was
done to obtain a better understanding of the overall shape of the $f_C(L_{\rm
2-8~keV})$ curve.  Figure~12a ({\it filled circles with error bars\/}) shows
our computed values of $\langle f_C \rangle_z$ as a function of $\log L_{\rm
2-8~keV}$.  We fit the $\langle f_C \rangle_z$ data points using a quadratic
relation (i.e., $\log \langle f_C \rangle_z = a_0 + a_1 \log L_{\rm 2-8~keV} +
a_2 (\log L_{\rm 2-8~keV})^2$, where $a_0=-136$, $a_1 = 7.1$, and $a_2 = -0.1$;
{\it thick solid curve} in Fig.~12a) over the luminosity range $\log L_{\rm
2-8~keV} = $~\hbox{40--44}; this covers the same luminosity range for AGNs as
the OS01 local sample.  Using our best-fit relation for $\langle f_C \rangle_z$
and our model for the redshift evolution of $f_C$, we calculated $f_C(z,\log
L_{\rm 2-8~keV})$ for each of our optically luminous faded samples following:

\begin{equation}
f_C=\left( \frac{1+z}{1+z_{\rm median}} \right )^3 \langle f_{C} \rangle_z,
\end{equation}

\noindent where $z$ is the median redshift of each sample, $z_{\rm
median}=0.55$ is the median redshift of our best-fit redshift-averaged
relation, $\langle f_C \rangle_z$.  In Figure~12a, we show our estimates of
$f_C$ for each of our optically luminous faded samples; these curves are
annotated on the figure.  For comparison, we have shown the AGN fraction
measured for DRGs (see $\S$~1) at $z \approx 2.5$ by Rubin \etal (2004; {\it
open diamond\/}) and have extrapolated our model for $f_C$ out to $z \approx
2.5$ ({\it dot-dashed curve\/}).  We note that even at $z \approx 2.5$ our
model agrees reasonably well with observations.

In order to estimate the amount of AGN contamination that may be contributing
to our stacked signals, it is desirable to convert the cumulative AGN fractions
to differential forms (i.e., the fraction of galaxies harboring AGNs within
discrete \xray\ luminosity bins).  Using the model for $f_C$ presented above,
we estimated the fractions of galaxies harboring an AGN within luminosity bins
of width $\Delta \log L_{\rm 2-8~keV} = 0.5$ for each of our optically luminous
faded samples; we refer to these as differential AGN fractions, $f_D$.  The
histograms in Figure~12b show our estimates of $f_D$.  Due to the deep \chandra\
coverage in the \ecdfs\ region, a large fraction of the luminous AGNs ($L_{\rm
2-8~keV} \simgt 10^{41.5}$~\xlum) would have been removed from our samples
before stacking (see $\S$~2.1.1).  We are therefore only interested in the
fraction of galaxies falling below our sensitivity limit.  Using the
sensitivity maps described in $\S$~3.1, we determined the fraction of optically
luminous galaxies within each of our stacked faded samples for which we could
{\it not} have detected an AGN of luminosity $L_{\rm 2-8~keV}$ if present; we
refer to these fractions as $f_B$ (i.e., the fractions of galaxies below our
sensitivity limit) and show them in Figure~12c.  For each of our optically
luminous faded samples we calculated the fraction of sources that harbor an
undetected AGN with \hbox{2--8~keV} luminosity $L_{\rm 2-8~keV}$ (in bins of
width $\Delta \log L_{\rm 2-8~keV} = 0.5$), $f_U$, by multiplying $f_D$ by
$f_B$.  In Figure~12d, we show $f_U$ as a function of $\log L_{\rm 2-8~keV}$.
Using $f_U$, we computed the approximate AGN contamination with the following
summation:

\begin{equation}
L_{\rm 2-8~keV}({\rm contam}) = \sum_i f_{U,i} \times L_{{\rm 2-8~keV},i},
\end{equation}

\noindent where the summation is over all bins of $\Delta \log L_{\rm 2-8~keV}
= 0.5$.  We find $L_{\rm 2-8~keV}(\rm contam) \approx 10^{39.5-40.3}$~\xlum\
for our samples.  For our samples with $z \simgt 0.45$, AGNs with $L_{\rm
2-8~keV} \simgt 10^{41}$~\xlum\ contribute $\simgt$70\% of the total $L_{\rm
2-8~keV}(\rm contam)$ estimate.  Furthermore, when extrapolating our model for
$f_C$ down to much lower values of $L_{\rm 2-8~keV}$, we find no significant
difference in our estimates of $L_{\rm 2-8~keV}(\rm contam)$.  We note that
this result is mildly dependent on our extrapolation of $f_C$ to values of
$L_{\rm 2-8~keV} \approx 10^{39-40.5}$~\xlum.  If there exists a large
population of radiatively-inefficient low-luminosity AGNs in early-type
galaxies that radiate within this \xray\ luminosity range (e.g.,
advection-dominated accretion flows [ADAFs]), then we may be underestimating
$L_{\rm 2-8~keV}(\rm contam)$.  However, \chandra\ observations of $\approx$50
early-type galaxies in the local universe have revealed that the majority of
the central supermassive black holes in these galaxies are radiating at
extremely low efficiencies and typically have observed $L_{\rm X} \simlt
10^{39}$~\xlum\ (e.g., Loewenstein \etal 2001; David \etal 2005;
Pellegrini~2005).  Therefore a change in the shape of $f_C$ at $L_{\rm 2-8~keV}
\approx 10^{39-40.5}$~\xlum\ is not expected.

We converted $L_{\rm 2-8~keV}(\rm contam)$ to a \hbox{0.5--2.0~keV} luminosity
assuming a power-law model with an effective photon index $\Gamma_{\rm eff}$,
which was determined by stacking all 24 AGN candidates in our optically
luminous faded sample; in this calculation we purposely made no attempt to
correct for intrinsic absorption.  For these 24 AGN candidates, we find that
the mean band ratio $\Phi_{\rm 2-8~keV}/\Phi_{\rm 0.5-2.0~keV}=1.12 \pm 0.09$,
which corresponds to an effective photon index of $\Gamma_{\rm eff} = 0.77 \pm
0.08$; this effective photon index is in agreement with the effective photon
indices measured for sources in the \chandra\ Deep Field-North (\hbox{CDF-N};
see Fig.~14 of Alexander \etal 2003) with similar fluxes.  If we assume
$\Gamma_{\rm eff} = 0.8$ is a characteristic effective photon index for the
AGNs we expect to be missing, then we find that AGN contamination can account
for $\approx$4\% and $\approx$11\% of the observed \hbox{0.5--2.0~keV} emission
from our optically luminous faded samples at $z \approx 0.25$ and $z \approx
0.66$, respectively; this amount does not significantly affect the results
presented in $\S$~3.2.1 above.  We note that the \xray\ SED used in this
calculation has an important effect on the overall estimate of the AGN
contamination.  Since our estimate for contamination decreases as $\Gamma_{\rm
eff}$ decreases, the amount of contamination in our samples may be affected if
our choice of $\Gamma_{\rm eff}$ is too flat.  However, if we choose a steeper
effective photon index such as $\Gamma_{\rm eff} = 1.4$ (the observed spectrum
of the \xray\ background), we still find that the estimated AGN contribution to
our \hbox{0.5--2.0~keV} signal is too low ($\simlt$25\% at $z \approx 0.66$) to
make a substantive difference to our results.  Furthermore, as discussed in
$\S$~2.1.1, we have taken additional precautions to eliminate several AGN
candidates that were not detected in the \hbox{2--8~keV} bandpass; these
sources would not be taken into account in this estimate for AGN contamination.
We also note that similar results are found when performing the above analyses
using our optically luminous general samples.

Due to poor statistical constraints on AGN activity in optically faint
early-type galaxies, an estimate of $f_C$ could not be determined reliably
using the present data.  Studies of the AGN fraction as a function of galactic
stellar mass have found that AGNs are much more commonly observed in massive
galaxies than lower-mass galaxies (e.g., Kauffmann \etal 2003).  We therefore
approximated a strict upper limit to the AGN contribution to the \xray\
emission from our optically faint early-type samples by using the same model
for $f_C$ presented above for the more massive optically luminous galaxies.
Using this model, we estimate that AGNs contribute $<$~20\% and $<$~40\% of the
\xray\ emission from our optically faint samples (both general and faded)
assuming $\Gamma_{\rm eff}=0.8$ and $\Gamma_{\rm eff} = 1.4$, respectively.
These limits slightly reduce the significance of the quoted evolution for our
optically faint faded samples such that $(L_{\rm X}/L_{B})_{z=0.5}$ is
estimated to be [$4.5 \pm 3.3$] and $[3.4 \pm 2.5] \times (L_{\rm
X}/L_{B})_{z=0}$ for $\Gamma_{\rm eff}=0.8$ and $\Gamma_{\rm eff} = 1.4$,
respectively; however, we note that these limits should be regarded as very
conservative.

\section{Discussion}

The above results suggest differing evolutionary histories for optically
luminous and faint early-type galaxies.  As shown in $\S$~3.2.2, our results
are not expected to be significantly affected by an undetected population of
AGNs.  Therefore, changes in the \xray\ emission with redshift are likely the
result of global changes in the emission from hot interstellar gas and/or
LMXBs.  In the sections below, we discuss possible interpretations of the
results for our optically luminous and faint early-type galaxies in turn.

\subsection{Optically Luminous Early-Type Galaxies}

As discussed in $\S$~2.2, we chose the SB for our stacking analyses to sample
directly the \xray\ emission from hot interstellar gas.  Therefore, the near
constancy of $L_{\rm X}/L_{B}$ with redshift can be largely explained as a
general balance between the energy losses from the hot gas ($\Delta E_{\rm
gas}$) and the energy deposition from heating mechanisms ($\Delta E_{\rm
heating}$) over each cooling time, $t_{\rm cool}$.  Here we investigate the
relative contributions from feedback mechanisms to constrain physical
models of the heating of the hot gas in early-type galaxies.  As noted in
$\S$~1.2, the typical inferred radiative cooling time for the central regions
of an optically luminous early-type galaxy is $t_{\rm cool} \approx 10^8$~yr.
Since each of our optically luminous early-type galaxy redshift bins are larger
than the cooling timescale (i.e., our redshift bins have temporal widths in the
range of \hbox{$\approx$0.5--4.4$ \times 10^9$~yr}), we can estimate the
redshift-dependent energy components following:

\begin{equation}\eqnum{7a}
\Delta E_{\rm gas} = L_{\rm gas} t_{\rm cool}
\end{equation}
\begin{equation}\eqnum{7b}
\Delta E_{\rm heating} = \epsilon_{\rm rad} \gamma_{\rm BC} L_{\rm 2-8~keV, AGN} t_{\rm cool} + L_{\rm mech, AGN} t_{\rm cool} + \Delta E_{\rm other}.
\end{equation}

\noindent Here, $L_{\rm gas}$ is the redshift-dependent globally-averaged power
output from the gaseous component of our optically luminous early-type
galaxies.  The first two terms of equation~7b represent AGN heating from both
radiative and mechanical feedback power, respectively.  The radiative feedback
power is represented as the product of the average \hbox{2--8~keV} AGN
luminosity per galaxy ($L_{\rm 2-8~keV, AGN}$), its bolometric correction
factor ($\gamma_{\rm BC} \approx 30$; e.g., Marconi \etal 2004; Barger \etal
2005), and the efficiency factor describing the coupling between radiation and
the hot interstellar gas ($\epsilon_{\rm rad}$).  We note that in cases where
the Compton temperature of the AGN SED falls below the temperature of the hot
interstellar gas, radiation from the central AGN may effectively cool the gas
and thereby drive $\epsilon_{\rm rad}$ to negative values (see, e.g., $\S$~6 of
Nulsen \& Fabian 2000; Ciotti \& Ostriker 2001).  Mechanical power (e.g.,
through AGN jets) is likely to be a very important feedback mechanism, and we
have indicated its contribution as $L_{\rm mech, AGN}$.  Finally, $\Delta
E_{\rm other}$ represents additional energy input from alternative forms of
heating over each cooling time (see below).  

Using the hot-gas component of the \xray\ SED for NGC~1600 (see {\it dotted
curve} in Fig.~1a), we estimate $L_{\rm gas} \approx 3 \times L_{\rm
0.5-2.0~keV}$; for our optically luminous samples, this amounts to a mean value
of $\langle L_{\rm gas} \rangle \approx 8 \times 10^{40}$~\xlum.  We measured
$L_{\rm 2-8~keV, AGN}$ using our redshift-dependent model for the differential
fraction, $f_D$, which was presented in $\S$~3.2.2 (see also Fig.~12b).  Duty
cycles for AGN activity in any given galaxy are expected to be shorter than the
timescales represented by each of our redshift bins; however, since we are
considering large populations of early-type galaxies, we do not expect
significant variations in the AGN fraction (as measured from the \hbox{E-CDF-S}
``snapshot'') at any given time within each redshift bin.  Therefore using this
model, we can calculate $L_{\rm 2-8~keV, AGN}$ following:

\begin{equation}\eqnum{8}
L_{\rm 2-8~keV, AGN} \approx  \sum_i f_{D,i} \times L_{{\rm 2-8~keV},i},
\end{equation}

\noindent where the summation is over bins of $\Delta \log L_{\rm 2-8~keV} =
0.5$ and covers the luminosity range $L_{\rm 2-8~keV}= 10^{40-44}$~\xlum.  We
found $L_{\rm 2-8~keV, AGN} \approx 9.1$ and $\approx$21~$\times
10^{40}$~\xlum\ per galaxy at $z \approx 0.25$ and $z \approx 0.66$,
respectively.  In Figure~13, we show $L_{\rm 2-8~keV, AGN}$ as a function of
redshift for our optically luminous faded samples ({\it filled squares\/}).
For comparison, we also show stacking results from the Brand \etal (2005)
samples of \hbox{$z \approx$~0.4--0.9} early-type galaxies ({\it open diamonds\/}),
which have mean $R$-band absolute magnitudes that are well matched to those of
our optically luminous faded samples; these mean luminosities are dominated by
\xray\ emission from AGNs and therefore provide a good estimate of $L_{\rm
2-8~keV, AGN}$ ($\approx$80--90\% of the emission is from AGNs).  Using these
data and our model for the evolution of the \xray\ emission from transient
AGNs, $L_{\rm 2-8~keV, AGN} = L_{\rm 2-8~keV, AGN, z=0} \; (1+z)^3$, we found
$L_{\rm 2-8~keV, AGN, z=0} \approx 5 \times 10^{40}$~\xlum\ (see the {\it
dotted curve} in Fig.~13).  

Using the above relations and the assumption that $\Delta E_{\rm gas} = \Delta
E_{\rm heating}$, we arrive at the following relation:

\begin{equation}\eqnum{9}
L_{\rm gas} = \epsilon_{rad} \gamma_{\rm BC} L_{\rm 2-8~keV, AGN, z=0} \; (1+z)^3 + L_{\rm mech, AGN} + \Delta E_{\rm other}/t_{\rm cool}
\end{equation}

\noindent As discussed in $\S$~1.2, transient AGN feedback is expected to play
a significant role in the heating of the hot interstellar gas.  If we assume
that AGN feedback is largely responsible for keeping the gas hot, then we can
neglect the last term of equation~9 (i.e., $\Delta E_{\rm other}/t_{\rm cool}
\ll L_{\rm gas}$).  With this assumption and the observed constancy of $L_{\rm
gas}$ with redshift, we infer that the strongly-evolving radiative power must
be poorly coupled to the hot interstellar gas, such that $\epsilon_{\rm rad}
\ll 0.05$.  This suggests that mechanical AGN power (i.e., $L_{\rm mech}$)
dominates the feedback over the redshift range $z \approx$ \hbox{0.0--0.7} and
does not evolve in the same way as the radiative power.  We note that it is
also possible that $\Delta E_{\rm other}$ may play some non-negligible role in
the heating of the hot gas.  Additional heating sources may include inward
thermal conduction from the large reservoirs of hot gas found in the outer
regions of early-type galaxies (e.g., Narayan \& Medvedev 2001; Brighenti \&
Mathews 2003), Type~Ia supernovae and stellar winds (e.g., Loewenstein \&
Mathews 1987, 1991), and infalling circumgalactic gas (e.g., Brighenti \&
Mathews 1998); however, the influence of these heating sources is presently not
well constrained.  

\subsection{Optically Faint Early-Type Galaxies}

For our optically faint early-type galaxy samples, we found suggestive evidence
for redshift evolution in $L_{\rm X}/L_B$ over the redshift range $z
\approx$~0.0--0.5 (see $\S$~3.2.1 and Figs.~10 and 11).  Although some of the
observed emission may be due to AGN activity ($\S$~3.2.2), there remains
suggestive evidence that normal \xray\ activity is evolving with redshift, and
we discuss possible scenarios explaining this evolution below.

One possible driver of $L_{\rm X}/L_B$ evolution may come from global changes
in the LMXB populations within optically faint early-type galaxies.  As
discussed in $\S$~1.2, LMXBs from primordial binaries within the galactic field
are expected to dominate the overall LMXB emission from optically faint
early-type galaxies; this differs from the LMXB emission from optically
luminous early-type galaxies, which originates primarily from globular
clusters.  LMXBs from primordial binaries emerge in the wake of star-formation
epochs $\approx$1--10~Gyr following a major star-formation event.  Therefore,
changes in the mean early-type galaxy stellar age with redshift should result
in observed changes in the mean \xray\ emission from these systems.
Furthermore, these changes are expected to be most conspicuous $\approx$1~Gyr
after major star-formation events (e.g., White \& Ghosh~1998; Ghosh \&
White~2001).  Galaxy formation scenarios that favor a more recent emergence of
the optically faint early-type galaxy population onto the red-sequence such as
downsizing or mass-dependent merging histories (see discussion and references
in $\S$~1) would predict significant evolution of the LMXB emission from these
systems.  

A second source of evolution of the \xray\ emission from optically faint
early-type galaxies could in principle come from cooling of the hot
\xray-emitting gas.  However, studies of optically faint early-type galaxies in
the local universe have shown that the hot gas emission generally makes up a
minority fraction (typically $\approx$40\%) of the total \xray\ emission (e.g.,
David \etal 2006) and is therefore less likely to be completely responsible for
the observed evolution than LMXBs.

\subsection{Future Work}

Understanding the evolution of the \xray\ properties of early-type
galaxies could be greatly improved by (1) constraining better the \xray\
properties of local optically faint early-type galaxies and making a census of
their AGN populations, (2) performing additional investigations using other
deep \chandra\ fields that have complementary \hst\ coverage, and (3)
conducting deeper observations using \chandra\ or future \xray\
missions.  These possibilities are discussed in more detail below.

In Figure~8, we showed the values of \Lx\ and $L_B$ for local early-type
galaxies from the OS01 sample, which is the largest, uniformly-selected sample
available for studying the \hbox{\Lx--$L_B$} correlations of local early-type
galaxies.  The majority ($\approx$75\%) of the isolated optically faint
early-type galaxies have only \xray\ upper limits, which has restricted our
interpretation of the redshift evolution of these galaxies.  New \chandra\
observations of the galaxies having only \xray\ upper limits could not only
improve the characterization of the \Lx--$L_B$ correlation at lower $L_B$, but
could also provide useful insight into the role of hot interstellar gas and
LMXBs within the galactic field, which are presently not well constrained.
Furthermore, stacking analyses of well-chosen samples of these galaxies could
provide useful statistical insight into the mean \xray\ properties of these
galaxies and mitigate the effects of poor source statistics for individual
galaxies.

\chandra\ stacking analyses using additional samples of early-type galaxies
could effectively reduce the sizes of the errors on mean quantities and/or
allow for the analyses of samples in more finely partitioned bins of redshift
and optical luminosity.  An important requirement for such studies is to obtain
an adequate census of the underlying AGN population, which may significantly
influence the stacking results.  In order to remove effectively AGNs from
stacked signals at $z \simlt 0.7$, relatively deep \chandra\ observations are
necessary.  Using the luminosity-dependent AGN fractions determined in
$\S$~3.2.2, we suggest that AGNs will provide significant contamination for
\chandra\ exposures of $\simlt$100~ks.  Therefore, studies of distant
early-type galaxies using multiwavelength data (including \hst\ coverage) from
already existing \chandra\ fields such as the $\approx$2~Ms \cdfn\ or the
$\approx$200~ks All-wavelength Extended Groth Strip International Survey
(AEGIS; Nandra \etal 2005; Davis \etal 2006) would improve the present
situation.

Finally, deeper \chandra\ observations of already existing fields (most notably
the \hbox{CDF-N}) would provide an improved census of the low-to-moderate luminosity
AGN population at higher redshifts and connect the \xray\ properties of these
relatively passive early-type populations with those of their higher redshift
progenitors (e.g., DRGs, EROs, and distant submillimeter-emitting galaxies;
e.g., Alexander \etal 2005).  In addition to the improvement that additional
\chandra\ observations could provide, it is also worth noting that future
\xray\ missions such as \xeus\ and \genx\footnote{For further information
regarding the future \xray\ missions \xeus\ and \genx, see
http://www.rssd.esa.int/XEUS/ and http://genx.cfa.harvard.edu, respectively.}
should allow the first investigations of the evolution of the normal early-type
galaxy \xray\ luminosity function with redshift.

\section{Summary}

Using \xray\ stacking analyses, we have investigated the \xray\ evolution of
539, $z \approx$~0.1--0.7 early-type galaxies located in the \ecdfs.  These
galaxies were selected using a combination of red-sequence colors and
S\'{e}rsic indices as a part of the COMBO-17 and GEMS surveys (M05).  We
classified our original early-type galaxy sample as the ``general sample'' and
generated an additional ``faded sample,'' which was corrected for the passive
fading of old stellar populations.  Using these samples, we analyzed separately
optically luminous ($L_B \approx 10^{10-11}$~\lbsol) and faint ($L_B \approx
10^{9.3-10}$~\lbsol) populations, which are expected to have soft \xray\
spectra dominated by hot interstellar gas and LMXBs, respectively. 
Our primary goal was to use stacking analyses to measure and constrain the
redshift evolution of the average \xray\ emission from normal early-type
galaxies.  To achieve this, we used a variety of techniques to identify
powerful AGNs, which we removed from our stacking analyses.  Our key results
are as follows:

\begin{enumerate}

\item We detected 49 early-type galaxies in the \xray\ band and classified 32
of these as AGN candidates based on their \xray, optical, and radio
properties (see $\S$~2.1.1 for details); the remaining 17 \xray-detected
sources had multiwavelength properties consistent with normal galaxies.  In
addition to the 32 \xray-detected AGN candidates, we identified 13 galaxies
with AGN-like radio--to--optical flux ratios, which we characterized as
potential AGNs.  We found that the majority of the AGN candidates were
coincident with optically luminous early-type hosts.  The inferred AGN fraction
for our optically luminous galaxies shows evidence for evolution
with redshift in a manner consistent with the $(1+z)^3$ evolution expected from
other studies of AGN evolution.  

\item When stacking the \xray\ counts from our normal optically luminous
early-type galaxy samples, we found that the \xray--to--optical mean luminosity
ratio, $L_{\rm X}/L_B$, stays roughly constant over the redshift range \hbox{$z
\approx$~0.0--0.7}, which indicates that the \xray-emitting gas has not
significantly evolved over the last $\approx$6.3~Gyr (i.e., since $z \approx
0.7$).  Using the data available, we found that $(L_{\rm X}/L_{B})_{z=0.7}$ =
$[1.0 \pm 0.5]$ and $[1.7 \pm 0.8] \times (L_{\rm X}/L_{B})_{z=0}$ for our
general and faded samples, respectively.  We interpret the lack of \xray\
evolution of optically luminous early-type galaxies to be due to an energy
balance between the heating and cooling of the hot gas over each cooling time.
When assuming that the heating is largely due to transient AGN activity, we
found that mechanical feedback dominates the heating out to $z \approx 0.7$
versus radiative power, which we inferred to be very poorly coupled to the gas.
Furthermore, this result suggests that the radiative and mechanical AGN power
evolve differently with cosmic time.

\item  For our optically faint early-type galaxy samples, we found suggestive
evidence that $L_{\rm X}/L_B$ increases with redshift.  By $z \approx$~0.5,
$L_{\rm X}/L_B$ is measured to be $5.3 \pm 4.1$ and $5.6 \pm 4.1$ times larger
than that measured at $z = 0$ for our general and faded samples, respectively;
however, due to poor statistical constraints on the local relation and the
undetected AGN population, we could not confidently rule out the null
hypothesis.  We hypothesized that evolution of the optically faint early-type
galaxy \xray\ emission may be due to the evolution of LMXBs in galaxies that
have recently joined the red-sequence and/or the cooling of hot gas within
these galaxies.

\end{enumerate}

\acknowledgements

We thank Ann Hornschemeier, Paul Nulsen, Ewan O'Sullivan, Craig Sarazin, Ohad
Shemmer, John Silverman, and the anonymous referee for useful suggestions,
which have improved the quality of this paper.  We gratefully acknowledge the
financial support of NSF CAREER award AST-9983783 (B.D.L., W.N.B.), \chandra\
X-ray Center grant G04-5157A (B.D.L., W.N.B., A.T.S.), the Royal Society
(D.M.A.), the Emmy Noether Program of the Deutsche Forschungsgemeinscaft
(E.F.B.), the \chandra\ Fellowship program (F.E.B.), and NSF grant AST 03-07582
(D.P.S.).  

%
{}
%
\clearpage
%
%

\begin{deluxetable}{ccccccccccc}
\tablenum{1}
\tabletypesize{\tiny}
\rotate
\tablewidth{0pt}
\tablecaption{X-ray Detected Early-Type Galaxies: Source Properties}

\tablehead{
\colhead{\chandra}                &
\colhead{}                        &
\multicolumn{2}{c}{Flux ($\log$ \flux)}  &
\colhead{Hardness Ratio}                &
\colhead{}                        &
\colhead{}                        &
\colhead{$L_B$}                   &
\colhead{$L_{\rm 2-8~keV}$}       &
\colhead{}                        &
\colhead{}                       \\

\colhead{Name (J2000.0)}          &
\colhead{$z$}                     &
\colhead{SB}                      &
\colhead{HB}                      &
\colhead{HB/SB}      &
\colhead{$\log f_{\rm 0.5-8.0~keV}/f_{R}$}         &
\colhead{$\Gamma$}                &
\colhead{($\log L_{B,\odot}$)}    &
\colhead{($\log$ \xlum)}          &
\colhead{Survey}                  &
\colhead{Notes}                  \\

\colhead{(1)}                     &
\colhead{(2)}                     &
\colhead{(3)}                     &
\colhead{(4)}                     &
\colhead{(5)}                     &
\colhead{(6)}                     &
\colhead{(7)}                     &
\colhead{(8)}                     &
\colhead{(9)}                     &
\colhead{(10)}                    &
\colhead{(11)}                    
}

\startdata
     J033121.17$-$275857.7 &                      0.68 &                   $-$15.5 &                   $-$14.7 &                      1.26 &                   $-$0.94 &                       0.7 &   11.1 &                         42.3 &           E-CDF-S 03 &                         A \\
     J033132.81$-$280115.9 &                      0.15 &                   $-$15.4 &               $<$ $-$14.9 &                  $<$ 0.78 &                   $-$2.63 &                       1.4 &   10.4 &                     $<$ 40.8 &           E-CDF-S 03 &                         N \\
     J033137.72$-$273843.3 &                      0.22 &               $<$ $-$15.5 &                   $-$14.6 &                  $>$ 1.46 &                   $-$1.93 &                       1.4 &   10.4 &                         41.4 &           E-CDF-S 02 &                         A \\
     J033138.05$-$280312.2 &                      0.49 &               $<$ $-$15.6 &               $<$ $-$14.9 &                   $\sim$1 &                   $-$1.04 &                       1.4 &   10.2 &                     $<$ 42.0 &           E-CDF-S 03 &                         A \\
     J033143.42$-$274248.6 &                      0.47 &                   $-$15.8 &                   $-$14.6 &                      2.31 &                   $-$0.77 &                       0.1 &   10.3 &                         42.0 &           E-CDF-S 02 &                         A \\
 \\
     J033151.15$-$275051.5 &                      0.68 &                   $-$15.3 &                   $-$14.9 &                      0.57 &                   $-$0.31 &                       1.2 &   10.3 &                         42.3 &                CDF-S &                       A,R \\
     J033156.00$-$273942.4 &                      0.58 &                   $-$15.2 &                   $-$14.9 &                      0.51 &                   $-$1.11 &                       1.5 &   10.8 &                         42.1 &           E-CDF-S 02 &                         A \\
     J033158.13$-$274459.4 &                      0.58 &                   $-$15.3 &                   $-$14.8 &                      0.66 &                   $-$0.83 &                       1.4 &   10.6 &                         42.2 &           E-CDF-S 02 &                         A \\
     J033200.42$-$275228.6 &                      0.63 &                   $-$15.9 &                   $-$15.6 &                      0.54 &                   $-$1.13 &                       1.4 &   10.4 &                         41.5 &                CDF-S &                         A \\
     J033200.83$-$275954.6 &                      0.43 &               $<$ $-$15.6 &                   $-$14.9 &                  $>$ 1.07 &                   $-$0.67 &                       1.4 &    9.6 &                         41.8 &           E-CDF-S 03 &                         A \\
 \\
     J033202.13$-$275621.6 &                      0.61 &               $<$ $-$15.6 &                   $-$14.8 &                  $>$ 1.49 &                   $-$0.54 &                       1.4 &   10.0 &                         42.2 &           E-CDF-S 03 &                         A \\
     J033203.65$-$274603.7 &                      0.59 &                   $-$15.1 &                   $-$13.7 &                      3.74 &                      0.01 &                    $-$0.3 &   10.8 &                         43.0 &           E-CDF-S 02 &                         A \\
     J033205.90$-$275449.7 &                      0.66 &                   $-$16.2 &               $<$ $-$15.0 &                  $<$ 3.18 &               $<$ $-$1.32 &                       1.4 &   10.9 &                     $<$ 42.1 &           E-CDF-S 03 &                       N,S \\
     J033206.27$-$274536.7 &                      0.66 &                   $-$15.9 &               $<$ $-$15.3 &                  $<$ 0.67 &                   $-$1.35 &                       1.4 &   10.7 &                     $<$ 41.8 &                CDF-S &                         N \\
     J033209.52$-$273634.1 &                      0.23 &                   $-$15.8 &               $<$ $-$14.9 &                  $<$ 1.85 &               $<$ $-$1.99 &                       1.4 &   10.2 &                     $<$ 41.2 &           E-CDF-S 02 &                         N \\
 \\
     J033214.36$-$274455.8 &                      0.57 &                   $-$15.5 &               $<$ $-$14.9 &                  $<$ 1.07 &                   $-$0.46 &                       1.4 &    9.8 &                     $<$ 42.1 &           E-CDF-S 02 &                         A \\
     J033217.06$-$274921.9 &                      0.34 &                   $-$16.3 &                   $-$15.3 &                      1.46 &                   $-$1.95 &                       0.4 &   10.4 &                         41.1 &                CDF-S &                         A \\
     J033218.44$-$274536.6 &                      0.47 &                   $-$16.4 &               $<$ $-$15.6 &                  $<$ 1.50 &               $<$ $-$2.06 &                       1.4 &   10.4 &                     $<$ 41.2 &                CDF-S &                       N,S \\
     J033218.45$-$274555.9 &                      0.69 &               $<$ $-$16.2 &               $<$ $-$15.4 &                   $\sim$1 &                   $-$1.66 &                       1.4 &   10.6 &                     $<$ 41.8 &                CDF-S &                       A,S \\
     J033220.48$-$274732.3 &                      0.60 &                   $-$16.2 &                   $-$14.1 &                     11.80 &                      0.34 &                    $-$1.4 &   10.1 &                         42.4 &                CDF-S &                         A \\
 \\
     J033221.99$-$274655.9 &                      0.64 &               $<$ $-$16.3 &                   $-$15.5 &                  $>$ 1.33 &                   $-$1.56 &                       1.4 &   10.6 &                         41.6 &                CDF-S &                         A \\
     J033224.26$-$274126.4 &                      0.48 &                   $-$14.7 &                   $-$14.3 &                      0.60 &                   $-$0.43 &                       1.3 &   10.5 &                         42.5 &           E-CDF-S 02 &                         A \\
     J033225.74$-$274936.4 &                      0.58 &                   $-$16.1 &               $<$ $-$15.2 &                  $<$ 1.37 &                   $-$1.90 &                       1.4 &   10.9 &                     $<$ 41.8 &                CDF-S &                         N \\
     J033228.81$-$274355.6 &                      0.22 &                   $-$15.2 &                   $-$15.1 &                      0.29 &                   $-$2.05 &                       1.9 &   10.5 &                         41.1 &                CDF-S &                       A,R \\
     J033229.22$-$274707.6 &                      0.66 &                   $-$16.4 &               $<$ $-$15.5 &                  $<$ 1.68 &                   $-$1.59 &                       1.4 &   10.6 &                     $<$ 41.7 &                CDF-S &                       N,S \\
 \\
     J033232.96$-$274545.7 &                      0.33 &                   $-$15.1 &               $<$ $-$14.7 &                  $<$ 0.56 &                   $-$1.19 &                       1.4 &   10.2 &                     $<$ 41.8 &           E-CDF-S 01 &                         N \\
     J033233.46$-$274312.8 &                      0.12 &                   $-$15.7 &                   $-$15.0 &                      0.90 &                   $-$2.74 &                       0.9 &   10.3 &                         40.4 &                CDF-S &                         A \\
     J033234.32$-$280018.0 &                      0.66 &               $<$ $-$15.5 &               $<$ $-$14.8 &                   $\sim$1 &                   $-$0.91 &                       1.4 &   10.5 &                     $<$ 42.4 &           E-CDF-S 01 &                       A,S \\
     J033234.34$-$274350.1 &                      0.63 &                   $-$15.8 &                   $-$15.2 &                      0.72 &                   $-$0.90 &                       1.0 &   10.5 &                         41.9 &                CDF-S &                         A \\
     J033237.31$-$274729.4 &                      0.64 &               $<$ $-$16.4 &                   $-$15.9 &                  $>$ 0.73 &               $<$ $-$2.01 &                       1.4 &   10.9 &                         41.2 &                CDF-S &                       A,S \\
 \\
     J033239.05$-$273456.4 &                      0.17 &                   $-$15.3 &               $<$ $-$14.8 &                  $<$ 0.69 &                   $-$0.22 &                       1.4 &    8.3 &                     $<$ 41.0 &           E-CDF-S 01 &                         A \\
     J033244.09$-$274541.5 &                      0.47 &                   $-$16.5 &               $<$ $-$15.5 &                  $<$ 1.90 &               $<$ $-$2.18 &                       1.4 &   10.7 &                     $<$ 41.3 &                CDF-S &                       N,S \\
     J033246.94$-$273902.8 &                      0.17 &                   $-$15.5 &               $<$ $-$14.9 &                  $<$ 1.05 &                   $-$2.87 &                       1.4 &   10.7 &                     $<$ 41.0 &           E-CDF-S 01 &                         N \\
     J033251.43$-$280304.4 &                      0.24 &                   $-$15.1 &               $<$ $-$14.9 &                  $<$ 0.44 &                   $-$2.49 &                       1.4 &   11.1 &                     $<$ 41.3 &           E-CDF-S 04 &                         N \\
     J033253.19$-$273902.4 &                      0.60 &                   $-$15.3 &                   $-$15.0 &                      0.51 &                   $-$0.42 &                       1.5 &   10.1 &                         42.0 &           E-CDF-S 01 &                         A \\
 \\
     J033256.33$-$274833.8 &                      0.11 &                   $-$15.7 &               $<$ $-$15.2 &                  $<$ 0.56 &                   $-$2.89 &                       1.4 &   10.2 &                     $<$ 40.2 &                CDF-S &                         N \\
     J033257.13$-$274534.3 &                      0.12 &                   $-$15.5 &               $<$ $-$15.0 &                  $<$ 0.86 &                   $-$2.54 &                       1.4 &   10.0 &                     $<$ 40.5 &           E-CDF-S 01 &                         N \\
     J033258.69$-$273738.3 &                      0.56 &               $<$ $-$15.6 &                   $-$15.3 &                  $>$ 0.45 &               $<$ $-$1.50 &                       1.4 &   10.8 &                         41.7 &           E-CDF-S 01 &                       A,S \\
     J033259.68$-$275030.3 &                      0.26 &                   $-$15.0 &               $<$ $-$14.8 &                  $<$ 0.39 &                   $-$1.31 &                       1.4 &   10.1 &                     $<$ 41.5 &           E-CDF-S 04 &                         N \\
     J033306.85$-$275448.7 &                      0.57 &                   $-$16.0 &               $<$ $-$15.3 &                  $<$ 1.24 &               $<$ $-$1.50 &                       1.4 &   10.8 &                     $<$ 41.7 &           E-CDF-S 04 &                         N \\
 \\
     J033311.55$-$275721.6 &                      0.16 &               $<$ $-$15.8 &                   $-$15.4 &                  $>$ 0.65 &               $<$ $-$1.31 &                       1.4 &    9.0 &                         40.4 &           E-CDF-S 01 &                     A,R,S \\
     J033312.63$-$275231.7 &                      0.64 &               $<$ $-$15.7 &                   $-$14.2 &                  $>$ 4.15 &                   $-$0.09 &                       1.4 &   10.4 &                         42.5 &           E-CDF-S 04 &                         A \\
     J033312.88$-$274219.8 &                      0.69 &               $<$ $-$15.8 &                   $-$14.6 &                  $>$ 2.60 &                   $-$0.50 &                       1.4 &   10.5 &                         42.2 &           E-CDF-S 01 &                         A \\
     J033319.58$-$274950.8 &                      0.65 &                   $-$14.9 &                   $-$13.9 &                      1.57 &                      0.46 &                       0.5 &   10.3 &                         43.0 &           E-CDF-S 04 &                         A \\
     J033320.60$-$274910.3 &                      0.14 &                   $-$14.6 &               $<$ $-$14.8 &                  $<$ 0.19 &                   $-$2.52 &                       1.4 &   10.8 &                     $<$ 40.9 &           E-CDF-S 01 &                         N \\
 \\
     J033320.85$-$274755.3 &                      0.14 &                   $-$14.4 &               $<$ $-$14.9 &                  $<$ 0.11 &                   $-$2.46 &                       1.4 &   10.9 &                     $<$ 40.8 &           E-CDF-S 01 &                         N \\
     J033324.22$-$273455.6 &                      0.51 &                   $-$14.9 &                   $-$14.5 &                      0.66 &                   $-$0.12 &                       1.3 &   10.1 &                         42.4 &           E-CDF-S 01 &                         A \\
     J033326.39$-$273521.8 &                      0.16 &               $<$ $-$15.6 &                   $-$14.4 &                  $>$ 2.40 &                   $-$2.02 &                       1.4 &   10.4 &                         41.3 &           E-CDF-S 01 &                         A \\
     J033328.86$-$273731.4 &                      0.36 &               $<$ $-$15.5 &               $<$ $-$14.9 &                   $\sim$1 &                   $-$1.97 &                       1.4 &   10.5 &                     $<$ 41.7 &           E-CDF-S 01 &                       A,S \\
\enddata
\tablecomments{Col.(1): \chandra\ source name. Col.(2): Source redshift as determined by \hbox{COMBO-17}. Col.(3)--(4): Flux for the \hbox{0.5--2.0~keV} and \hbox{2--8~keV} bandpasses. Col.(5): Hardness ratio of the \hbox{2--8~keV} and \hbox{0.5--2.0~keV} count rates ($\Phi_{2-8~keV}/\Phi_{0.5-2.0~keV}$). Col.(6): Logarithm of the \hbox{0.5--8.0~keV} to $R$-band flux ratio. Col.(7): Effective photon index ($\Gamma$). Here a value of 1.4 was assumed when photon statistics were too limited to determine accurate values. Col.(8): Logarithm of the rest-frame $B$-band luminosity. Col.(9): Logarithm of the rest-frame \hbox{2--8~keV} luminosity. Col.(10): Survey field in which each source was identified.  For E-CDF-S identifications, the associated field number (i.e., 01--04) indicates the \chandra\ pointing within which the source was detected (see Lehmer et~al. 2005b for details). Col.(11): Source notes.  Here an ``N'' denotes normal galaxies, an ``A'' denotes candidate AGNs, an ``R'' denotes sources with radio detections, and an ``S'' denotes sources that were detected in the supplementary catalogs (i.e., using {\ttfamily wavdetect} at a false-positive probability threshold of $1 \times 10^{-5}$).}
\end{deluxetable}

%
%

\begin{deluxetable}{cccccccccc}
\tablenum{2}
\tabletypesize{\tiny}
\rotate
\tablewidth{0pt}
\tablecaption{Stacked Early-Type Normal Galaxies: Basic Properties}

\tablehead{
\colhead{}                                &
\colhead{}                                &
\colhead{}                                &
\colhead{$E_{\rm 0.5-2.0~keV}$}           &
\multicolumn{3}{c}{Net Counts ($S-B$)}    &
\multicolumn{3}{c}{Signal-to-Noise Ratio (S/N)} \\

\colhead{$z_{\rm mean}$}                   &
\colhead{$N_{\rm total}$}                             &
\colhead{$N_{\rm detected}$}                             &
\colhead{(Ms)}           &
\colhead{0.5--1.0~keV}                    &
\colhead{0.5--2.0~keV}                    &
\colhead{2--8~keV}                        &
\colhead{0.5--1.0~keV}                    &
\colhead{0.5--2.0~keV}                    &
\colhead{2--8~keV}                        \\

\colhead{(1)}                             &
\colhead{(2)}                             &
\colhead{(3)}                             &
\colhead{(4)}                             &
\colhead{(5)}                             &
\colhead{(6)}                             &
\colhead{(7)}                             &
\colhead{(8)}                             &
\colhead{(9)}                             &
\colhead{(10)}                             
}

\startdata
\multicolumn{9}{c}{General Sample; $L_B \approx 10^{10-11}$~\lbsol} \\
\vspace{-0.1in} \\
\tableline 
 0.25 $\pm$ 0.08 &   45 &    5 &  15.0 &       30.5 $\pm$ 7.4 &      94.5 $\pm$ 11.9 &             $<$ 27.0 &                  9.8 &                 19.7 &                  0.9 \\ 
 0.47 $\pm$ 0.03 &   52 &    2 &  20.4 &       15.6 $\pm$ 6.6 &       31.4 $\pm$ 9.1 &             $<$ 32.4 &                  4.1 &                  5.5 &                  2.1 \\ 
 0.58 $\pm$ 0.02 &   51 &    2 &  18.2 &       12.9 $\pm$ 6.2 &       33.5 $\pm$ 9.0 &             $<$ 31.2 &                  3.6 &                  6.2 &                  2.2 \\ 
 0.66 $\pm$ 0.02 &   74 &    3 &  30.0 &       16.6 $\pm$ 7.3 &      40.7 $\pm$ 10.6 &      34.4 $\pm$ 13.4 &                  3.5 &                  5.7 &                  3.2 \\ 
\tableline \\
\vspace{-0.2in} \\
\multicolumn{9}{c}{General Sample; $L_B \approx 10^{9.3-10}$~\lbsol} \\
\vspace{-0.1in} \\
\tableline 
 0.24 $\pm$ 0.11 &   27 &    1 &  11.9 &       10.1 $\pm$ 5.4 &       25.5 $\pm$ 7.8 &             $<$ 27.1 &                  3.4 &                  5.7 &                  2.9 \\ 
 0.46 $\pm$ 0.03 &   27 &    0 &   9.9 &             $<$ 14.5 &       12.7 $\pm$ 6.5 &             $<$ 22.0 &                  2.7 &                  3.2 &                  0.5 \\ 
\tableline \\
\vspace{-0.2in} \\
\multicolumn{9}{c}{Faded Sample; $L_{B,0} \approx 10^{10-11}$~\lbsol} \\
\vspace{-0.1in} \\
\tableline 
 0.25 $\pm$ 0.08 &   41 &    4 &  14.1 &       26.3 $\pm$ 7.0 &      75.3 $\pm$ 10.9 &             $<$ 26.2 &                  8.7 &                 16.2 &                  1.0 \\ 
 0.47 $\pm$ 0.03 &   44 &    2 &  16.5 &             $<$ 17.0 &       25.4 $\pm$ 8.3 &             $<$ 30.0 &                  2.9 &                  4.9 &                  2.3 \\ 
 0.58 $\pm$ 0.02 &   30 &    2 &   9.5 &       12.5 $\pm$ 5.5 &       28.0 $\pm$ 7.6 &       21.9 $\pm$ 8.8 &                  4.9 &                  7.2 &                  3.6 \\ 
 0.66 $\pm$ 0.02 &   55 &    3 &  20.5 &       17.6 $\pm$ 6.7 &       42.2 $\pm$ 9.8 &             $<$ 33.5 &                  4.6 &                  7.3 &                  2.9 \\ 
\tableline \\
\vspace{-0.2in} \\
\multicolumn{9}{c}{Faded Sample; $L_{B,0} \approx 10^{9.3-10}$~\lbsol} \\
\vspace{-0.1in} \\
\tableline 
 0.22 $\pm$ 0.09 &   28 &    2 &  11.2 &       14.3 $\pm$ 5.8 &       43.5 $\pm$ 8.9 &             $<$ 26.3 &                  4.9 &                 10.1 &                  2.6 \\ 
 0.46 $\pm$ 0.03 &   31 &    0 &  12.5 &       11.9 $\pm$ 5.7 &       15.3 $\pm$ 7.1 &             $<$ 24.2 &                  3.9 &                  3.4 &                  0.3 \\ 
\enddata
\tablecomments{Col.(1): Mean redshift and standard deviation of the redshift for each stacked sample. Col.(2): Number of sources being stacked. Col.(3): Number of stacked sources that were detected individually in the \xray\ bandpass. Col.(4): Total vignetting-corrected effective exposure time measured from the \hbox{0.5--2.0~keV} exposure maps. Col.(5)--(7): Net \hbox{0.5--1.0~keV}, \hbox{0.5--2.0~keV}, and \hbox{2--8~keV} source counts. Col.(8)--(10): S/N for the \hbox{0.5--1.0~keV}, \hbox{0.5--2.0~keV}, and \hbox{2--8~keV} bandpasses.}
\end{deluxetable}

%
%
\clearpage
\thispagestyle{empty}
\begin{deluxetable}{cccccccccc}
\tablenum{3}
\tabletypesize{\tiny}
\rotate
\tablewidth{0pt}
\tablecaption{Stacked Early-Type Normal Galaxies: Mean X-ray Properties}

\tablehead{
\colhead{}                                       &
\colhead{$f_{\rm 0.5-1.0~keV}$}                  &
\colhead{$f_{\rm 0.5-2.0~keV}$}                  &
\colhead{$f_{\rm 2-8~keV}$}                      &
\colhead{}                                       &
\colhead{$L_{\rm 0.5-1.0~keV}$}                  &
\colhead{$L_{\rm 0.5-2.0~keV}$}                  &
\colhead{$L_{\rm 2-8~keV}$}                      &
\colhead{$L_B^a$}                                &
\colhead{$L_{\rm 0.5-2.0~keV}/L_B^a$}            \\

\colhead{$z_{\rm mean}$}                          &
\colhead{($\log$ cgs)}                           &
\colhead{($\log$ cgs)}                           &
\colhead{($\log$ cgs)}                           &
\colhead{$\log f_{\rm 0.5-8.0~keV}/f_R$}         &
\colhead{($\log$~\xlum)}                         &
\colhead{($\log$~\xlum)}                         &
\colhead{($\log$~\xlum)}                         &
\colhead{($\log$~\lbsol)}                        &
\colhead{($\log$~ergs~s$^{-1}$~\lbsol$^{-1}$)}   \\

\colhead{(1)}                             &
\colhead{(2)}                             &
\colhead{(3)}                             &
\colhead{(4)}                             &
\colhead{(5)}                             &
\colhead{(6)}                             &
\colhead{(7)}                             &
\colhead{(8)}                             &
\colhead{(9)}                             &
\colhead{(10)}                             
}

\startdata
\multicolumn{10}{c}{General Sample; $L_B \approx 10^{10-11}$~\lbsol} \\
\vspace{-0.1in} \\
\tableline 
 0.25 $\pm$ 0.08 &    $-$16.5 $\pm$ 0.6 &    $-$16.2 $\pm$ 0.2 &          $<$ $-$16.0 &              $-$3.00 &       39.7 $\pm$ 0.6 &       40.2 $\pm$ 0.2 &             $<$ 40.4 &                 10.5 &       29.7 $\pm$ 0.1 \\ 
 0.47 $\pm$ 0.03 &    $-$16.9 $\pm$ 0.7 &    $-$16.8 $\pm$ 0.4 &          $<$ $-$16.1 &              $-$2.63 &       39.9 $\pm$ 0.8 &       40.1 $\pm$ 0.4 &             $<$ 40.8 &                 10.4 &       29.7 $\pm$ 0.1 \\ 
 0.58 $\pm$ 0.02 &    $-$17.0 $\pm$ 0.8 &    $-$16.8 $\pm$ 0.3 &          $<$ $-$16.1 &              $-$2.29 &       40.1 $\pm$ 0.8 &       40.4 $\pm$ 0.4 &             $<$ 41.1 &                 10.5 &       29.9 $\pm$ 0.1 \\ 
 0.66 $\pm$ 0.02 &    $-$17.1 $\pm$ 0.8 &    $-$16.9 $\pm$ 0.3 &    $-$16.4 $\pm$ 0.5 &              $-$2.21 &       40.1 $\pm$ 0.8 &       40.4 $\pm$ 0.4 &       40.8 $\pm$ 0.6 &                 10.6 &       29.9 $\pm$ 0.1 \\ 
\tableline \\
\vspace{-0.2in} \\
\multicolumn{10}{c}{General Sample; $L_B \approx 10^{9.3-10}$~\lbsol} \\
\vspace{-0.1in} \\
\tableline 
 0.24 $\pm$ 0.11 &    $-$17.1 $\pm$ 0.9 &    $-$16.7 $\pm$ 0.4 &          $<$ $-$15.9 &              $-$2.70 &       39.4 $\pm$ 0.9 &       39.6 $\pm$ 0.4 &             $<$ 40.5 &                  9.8 &       29.9 $\pm$ 0.2 \\ 
 0.46 $\pm$ 0.03 &          $<$ $-$16.8 &    $-$17.0 $\pm$ 0.7 &          $<$ $-$16.0 &            $< -$1.79 &             $<$ 40.5 &       40.0 $\pm$ 0.7 &             $<$ 41.0 &                  9.6 &       30.4 $\pm$ 0.2 \\ 
\tableline \\
\vspace{-0.2in} \\
\multicolumn{10}{c}{Faded Sample; $L_{B,0} \approx 10^{10-11}$~\lbsol} \\
\vspace{-0.1in} \\
\tableline 
 0.25 $\pm$ 0.08 &    $-$16.5 $\pm$ 0.6 &    $-$16.3 $\pm$ 0.2 &          $<$ $-$16.0 &              $-$3.08 &       39.7 $\pm$ 0.6 &       40.1 $\pm$ 0.2 &             $<$ 40.4 &                 10.4 &       29.7 $\pm$ 0.1 \\ 
 0.47 $\pm$ 0.03 &          $<$ $-$16.6 &    $-$16.8 $\pm$ 0.4 &          $<$ $-$16.1 &              $-$2.67 &             $<$ 40.2 &       40.1 $\pm$ 0.4 &             $<$ 40.9 &                 10.4 &       29.8 $\pm$ 0.2 \\ 
 0.58 $\pm$ 0.02 &    $-$16.7 $\pm$ 0.7 &    $-$16.5 $\pm$ 0.3 &    $-$16.1 $\pm$ 0.5 &              $-$2.21 &       40.3 $\pm$ 0.8 &       40.6 $\pm$ 0.4 &       41.0 $\pm$ 0.6 &                 10.5 &       30.2 $\pm$ 0.1 \\ 
 0.66 $\pm$ 0.02 &    $-$16.9 $\pm$ 0.7 &    $-$16.7 $\pm$ 0.3 &          $<$ $-$16.1 &              $-$2.22 &       40.3 $\pm$ 0.7 &       40.6 $\pm$ 0.4 &             $<$ 41.2 &                 10.5 &       30.1 $\pm$ 0.1 \\ 
\tableline \\
\vspace{-0.2in} \\
\multicolumn{10}{c}{Faded Sample; $L_{B,0} \approx 10^{9.3-10}$~\lbsol} \\
\vspace{-0.1in} \\
\tableline 
 0.22 $\pm$ 0.09 &    $-$17.0 $\pm$ 0.7 &    $-$16.4 $\pm$ 0.3 &          $<$ $-$15.9 &              $-$2.71 &       39.5 $\pm$ 0.7 &       39.8 $\pm$ 0.3 &             $<$ 40.4 &                  9.8 &       30.0 $\pm$ 0.1 \\ 
 0.46 $\pm$ 0.03 &    $-$17.1 $\pm$ 0.8 &    $-$17.0 $\pm$ 0.6 &          $<$ $-$16.0 &            $< -$1.96 &       40.1 $\pm$ 0.8 &       40.0 $\pm$ 0.6 &             $<$ 40.9 &                  9.8 &       30.2 $\pm$ 0.2 \\ 
\enddata
\tablenotetext{a}{For the faded samples, quoted $B$-band luminosities indicate faded ($z=0$) luminosities, $L_{B,0}$.}
\tablecomments{Col.(1): Mean redshift and standard deviation of the redshift for each stacked sample. Col.(2)--(4): Logarithm of the mean \hbox{0.5--1.0~keV}, \hbox{0.5--2.0~keV}, and \hbox{2--8~keV} flux in units of \flux. Col.(5): Logarithm of the \hbox{0.5--8.0~keV} to $R$-band flux ratio. Col.(6)--(8): Logarithm of the mean \hbox{0.5--1.0~keV}, \hbox{0.5--2.0~keV}, and \hbox{2--8~keV} rest-frame luminosity. Col.(9): Logarithm of the mean rest-frame $B$-band luminosity. Col.(10): Logarithm of the \hbox{0.5--2.0~keV}--to--$B$-band mean luminosity ratio.}
\end{deluxetable}

\clearpage
%
%

\begin{figure}
\figurenum{1}
\centerline{
\includegraphics[width=14.0cm]{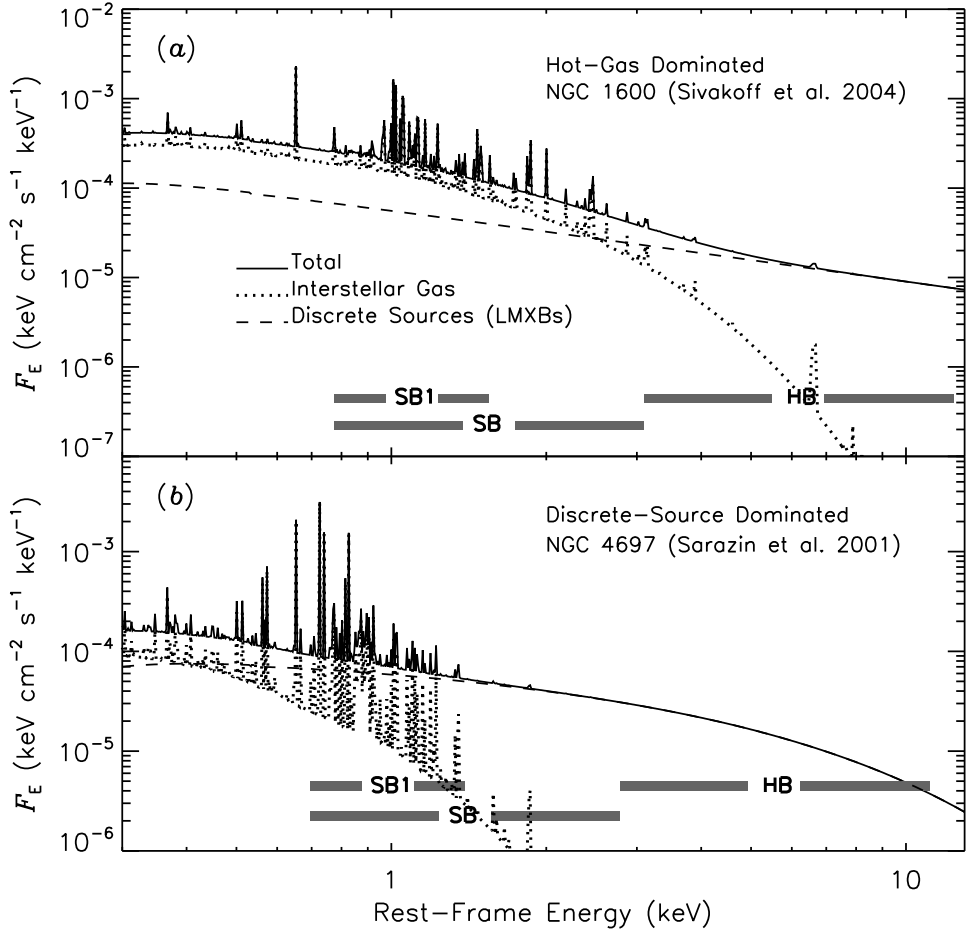}
}
\caption{
\footnotesize
Rest-frame X-ray spectral energy distributions (SEDs) for ({\bf a}) NGC~1600
(Sivakoff \etal 2004) and ({\bf b}) NGC~4697 (Sarazin \etal 2001).  These are
normal early-type galaxies with \xray\ emission dominated by hot interstellar
gas (NGC~1600) and discrete LMXBs (NGC~4697) and have \xray\ spectra
representative of optically luminous and faint galaxies, respectively.  These
SEDs ({\it solid curves\/}) were separated into LMXB ({\it dashed curves\/}) and
hot gas ({\it dotted curves\/}) components (see $\S$~2.2 for further details).
The \xray\ emission from NGC~1600 was best fit using a power-law ($\Gamma
\approx 1.8$) for the discrete component and a {\ttfamily MEKAL} plasma ($kT
\approx 1$~keV; $Z \approx 0.2 Z_{\odot}$) for the unresolved component.
NGC~4697 was best fit by a thermal bremsstrahlung ($kT \approx 5.2$~keV) for
the LMXB component and a {\ttfamily MEKAL} plasma ($kT \approx 0.3$~keV; $Z
\approx 0.1 Z_{\odot}$) for the hot interstellar gas.  For reference, we have
shown the energy ranges of our adopted bandpasses redshifted to $z=0.55$ and
$z=0.39$, the median redshifts of our optically luminous and faint samples,
respectively.  We note that for \xray\ luminous (optically luminous) early-type
galaxies within the redshift ranges considered in this study, the
\hbox{0.5--1.0~keV} (SB1) and \hbox{0.5--2.0~keV} (SB) bandpasses primarily
trace hot interstellar gas and the \hbox{2--8~keV} (HB) bandpass is primarily
dominated by LMXBs.  However, for \xray\ faint (optically faint) early-type
galaxies, all bandpasses are generally dominated by LMXB emission.}
\end{figure}

%
%

\begin{figure}
\figurenum{2}
\centerline{
\includegraphics[width=14cm]{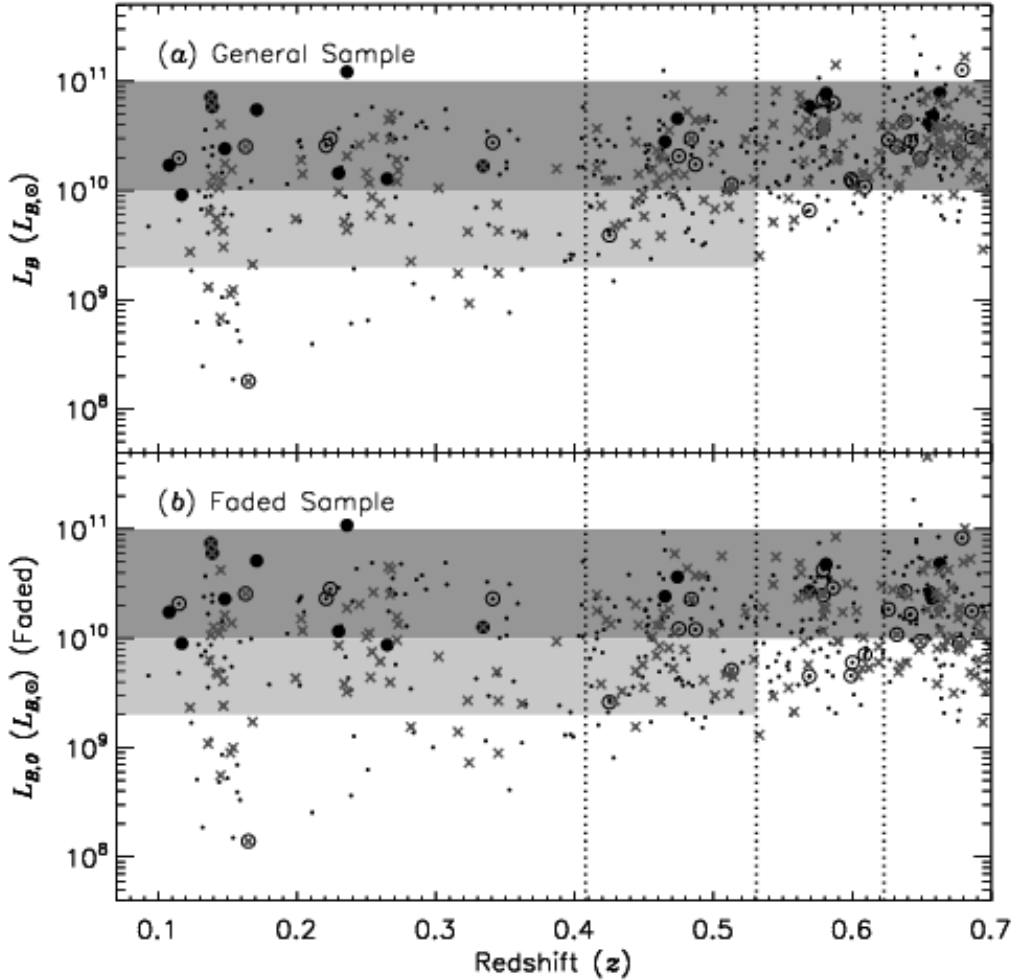}
}
\caption{
\footnotesize
({\bf a}) Rest-frame $B$-band luminosity versus redshift for our general
sample of 539 early-type galaxies.  Larger circles indicate \xray-detected AGN
candidates ({\it open\/}) and normal galaxies ({\it filled\/}).  Sources
denoted with crosses are (1) in close proximity ($<$10\arcsec) to an
\xray-detected source, (2) within the boundaries of extended \xray\ sources,
(3) at large off-axis angles (i.e., $>$7\arcmin\ from all aimpoints), and/or
(4) found to have relatively large radio--to--optical flux ratios (see
$\S$~2.1.1); these sources have been removed from our stacking analyses (see
$\S$~2.2 for details).  The shaded bands show the luminosity ranges of our
optically luminous and faint samples; vertical dotted lines indicate the
evenly-spaced comoving volume intervals chosen in constructing our stacking
samples.  ({\bf b}) Evolved, $z=0$, $B$-band luminosity of our sample (see
discussion in $\S$~2.1.2), which constitutes our faded sample.  Symbols and
boundaries are the same as in Figure~3a.  Using a Chabrier initial mass
function (Chabrier 2003), we estimate that the luminosity ranges $L_{B,0} =
10^{9.3-10}$ and $10^{10-11}$~\lbsol\ correspond roughly to stellar mass ranges
of $\approx 10^{9.9-10.6}$ and $\approx 10^{10.6-11.6}$~\msol, respectively.
} \end{figure}

%
%

\begin{figure}
\figurenum{3}
\centerline{
\includegraphics[width=16cm]{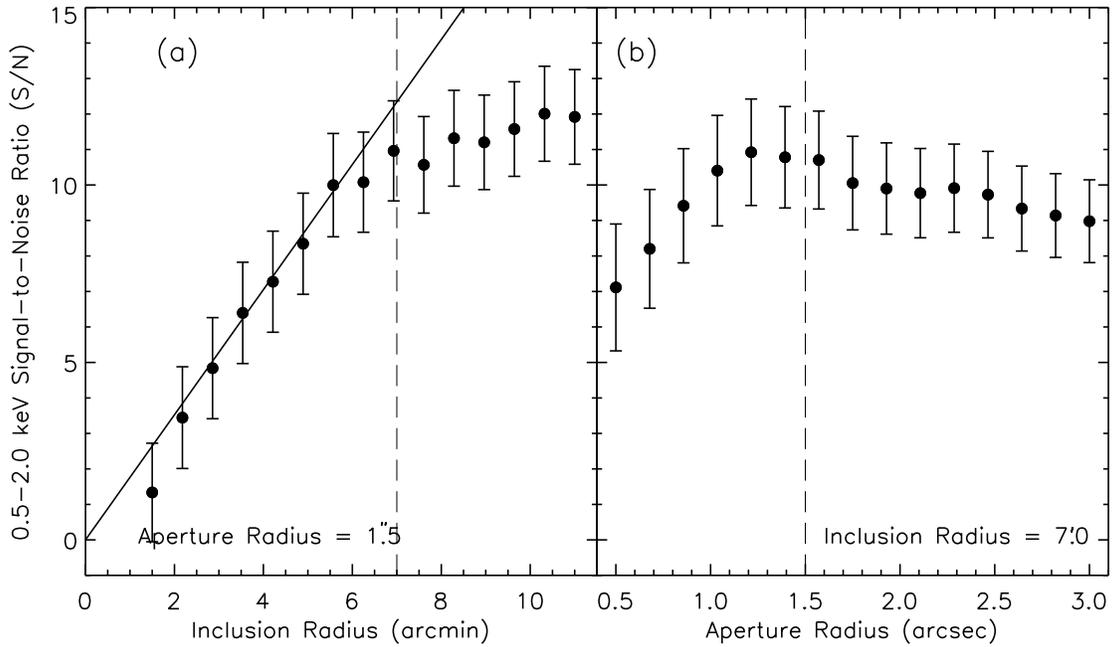}
}
\caption{
\footnotesize
({\bf a}) Soft-band (0.5--2.0~keV) signal-to-noise ratio (S/N) versus inclusion
radius (i.e., the maximum off-axis angle within which a source is included in
stacking) for a fixed aperture size of 1\farcs5; the solid line indicates the
expected linear increase in S/N versus inclusion radius.  ({\bf b}) Soft-band
S/N versus aperture radius for a fixed inclusion radius of 7\farcm0.  The above
plots were generated using corresponding optimized stacking parameters (i.e.,
stacking aperture size and inclusion radius), which were determined iteratively
(see $\S$~2.2 for details); the optimized parameters are indicated by vertical
dashed lines in each plot. } \end{figure}

%
%

\begin{figure}
\figurenum{4}
\centerline{
\includegraphics[width=14cm]{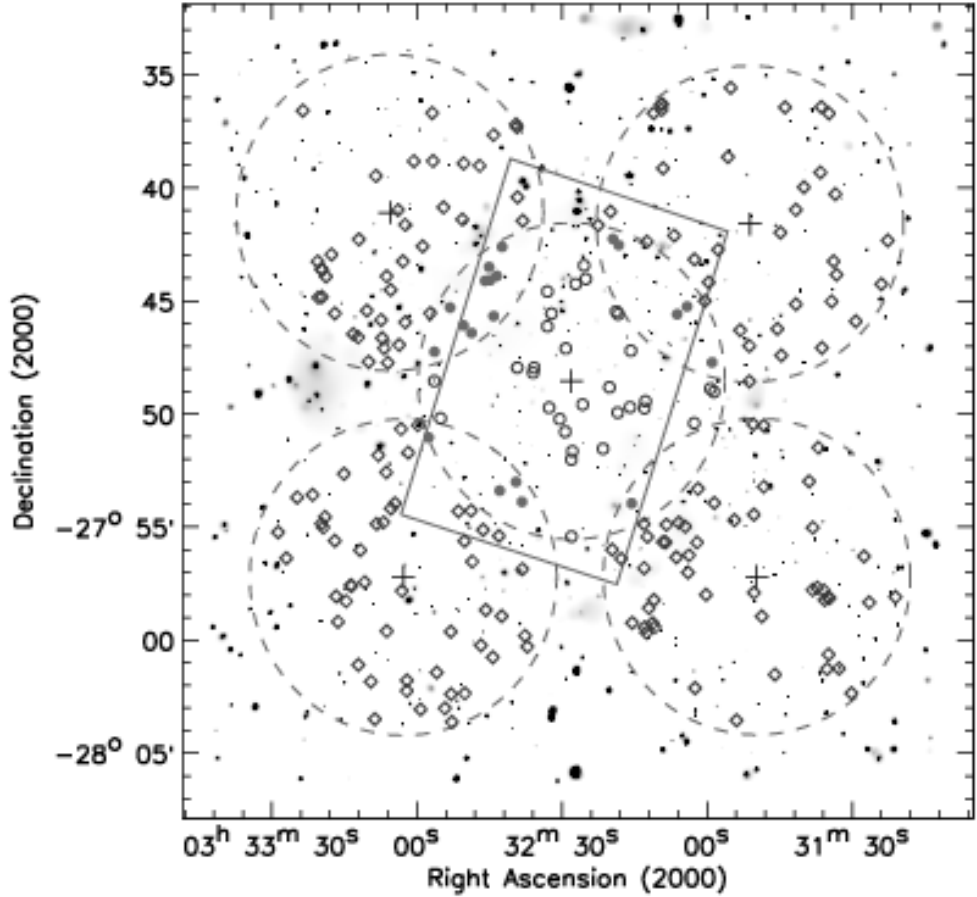}
}
\caption{
\footnotesize
Adaptively smoothed \hbox{0.5--2.0~keV} image of the combined $\approx$1~Ms
\cdfs\ and $\approx$250~ks \ecdfs.  Positions of stacked sources from our faded
sample are shown as open circles ($\approx$1~Ms \cdfs) and open diamonds
($\approx$250~ks \ecdfs); filled circles represent galaxy positions that have
been stacked using both the $\approx$1~Ms \cdfs\ and $\approx$250~ks \ecdfs\
observations (see $\S$~2.2 for additional details).  Aim points of each
\chandra\ observation are indicated as plus signs, and the surrounding
7\farcm0 inclusion radii are indicated with dashed circles.  The apparent lack
of sources in the north-eastern corner (i.e., the upper left-hand corner) of
the image is partially due to missing \hst\ coverage from the GEMS imaging.
For reference, we have outlined the $\approx$160~arcmin$^2$ GOODS-S region
({\it solid rotated rectangle}; Giavalisco \etal 2004).} \end{figure}

%
%

\begin{figure}
\figurenum{5}
\centerline{
\includegraphics[width=14cm]{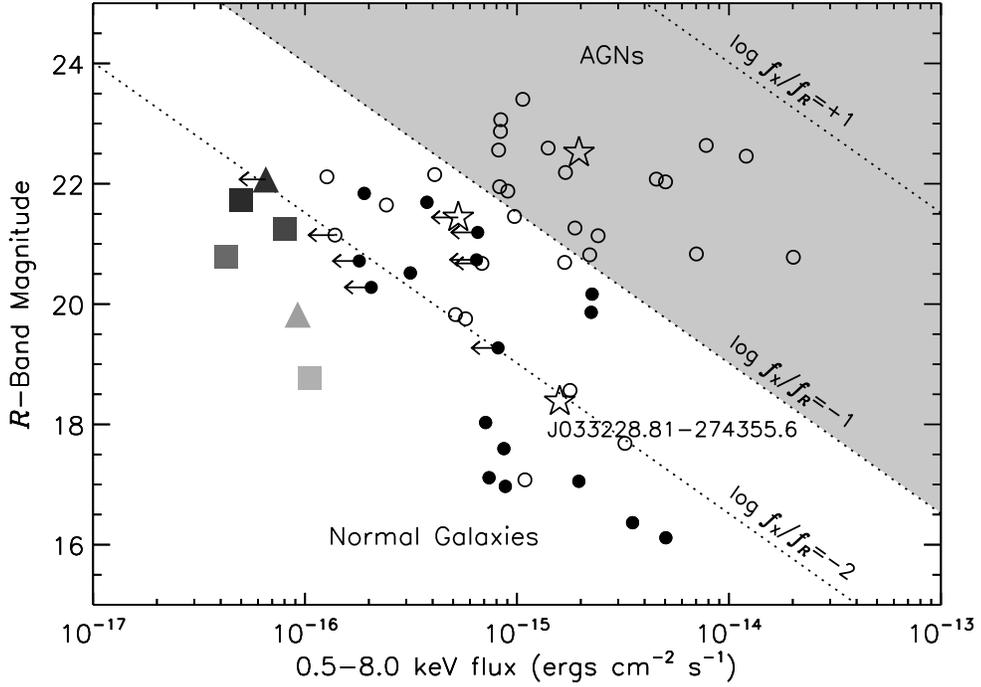}
}
\caption{
\footnotesize
Optical $R$-band magnitude versus \hbox{0.5--8.0~keV} flux for
\xray-detected early-type galaxies in our sample.  AGN candidates and normal
galaxies are plotted as open symbols ({\it circles and stars\/}) and filled
circles, respectively; sources with upper limits were detected in either the
\hbox{0.5--2.0~keV} or \hbox{2--8~keV} bandpasses.  Diagonal dotted lines
represent lines of constant \xray--to--optical flux ratio (i.e., $\log f_{\rm
0.5-8.0~keV}/f_R = $+1, $-$1, and $-$2); luminous AGNs generally have $\log
f_{\rm 0.5-8.0~keV}/f_R \simgt -1$ ({\it shaded region\/}).  AGN-candidates
were classified following the three criteria discussed in $\S$~2.1.1.  The
sources marked with five-pointed stars are radio-detected AGN candidates,
including the FR~II source CXOECDFS \hbox{J033228.81--274355.6}.  Furthermore,
mean $R$-band magnitudes and \hbox{0.5--8.0~keV} fluxes for our stacked
optically luminous ({\it large filled squares\/}) and faint ({\it large filled
triangles\/}) faded samples are plotted with varying grayscale levels to
indicate redshift; darker levels indicate larger redshift values.  }
\end{figure}

%
%

\begin{figure}
\figurenum{6}
\centerline{
\includegraphics[width=14cm]{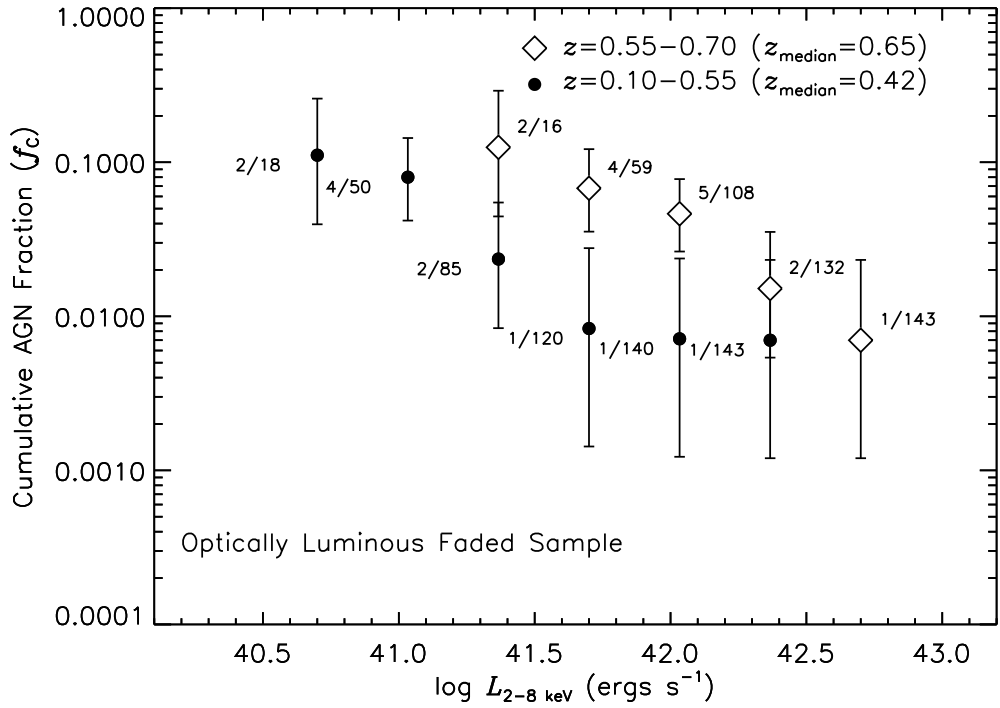}
}
\caption{
\footnotesize
Cumulative \xray\ detected AGN fraction, $f_C$, as a function of
\hbox{2--8~keV} luminosity for our optically luminous faded samples with $z
\approx$~0.10--0.55 ({\it filled circles\/}) and \hbox{$z \approx$~0.55--0.70}
({\it open diamonds\/}).  Each data point with printed fractions represents the
number of AGN candidates detected with \hbox{2--8~keV} luminosity of $L_{\rm
2-8~keV}$ or greater divided by the number of early-type galaxies with \xray\
coverage sufficient to detect an AGN of $L_{\rm 2-8~keV}$ (see $\S$~3.1 for
further details).} \end{figure}

%
%

\begin{figure}
\figurenum{7}
\centerline{
\includegraphics[width=18cm]{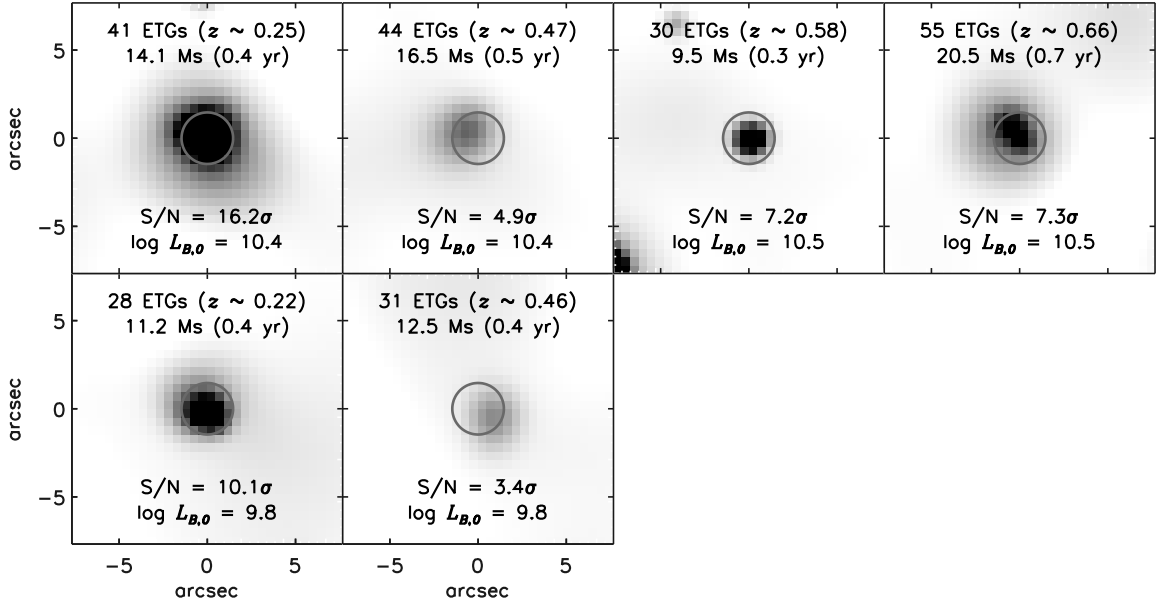}
}
\caption{
\footnotesize
Stacked, adaptively-smoothed \hbox{0.5--2.0~keV} images of our
early-type, faded samples for each redshift bin.  The top panels show our
optically luminous ($L_{B,0} \approx 10^{10-11}$~\lbsol) samples, and the
bottom panels show our optically faint ($L_{B,0} \approx 10^{9.3-10}$~\lbsol)
samples.  These images were generated using the {\ttfamily CIAO} tool {\ttfamily
CSMOOTH} with a minimum significance of 2.5$\sigma$.  The images are
$\approx$15\arcsec\ ($\approx$30.5 pixels) per side, and each pixel is
0\farcs492.  Our circular stacking aperture of radius 1\farcs5 is shown in each
image centered on the optical centroid of our stacked sources.  Additional
sample information, including the number of stacked galaxies, total \chandra\
exposure, signal-to-noise ratio (S/N), and logarithm of the mean $z=0$ $B$-band
luminosity ($L_{B,0}$) are annotated on each smoothed image.  } \end{figure}

%
%

\begin{figure}
\figurenum{8}
\centerline{
\includegraphics[width=14cm]{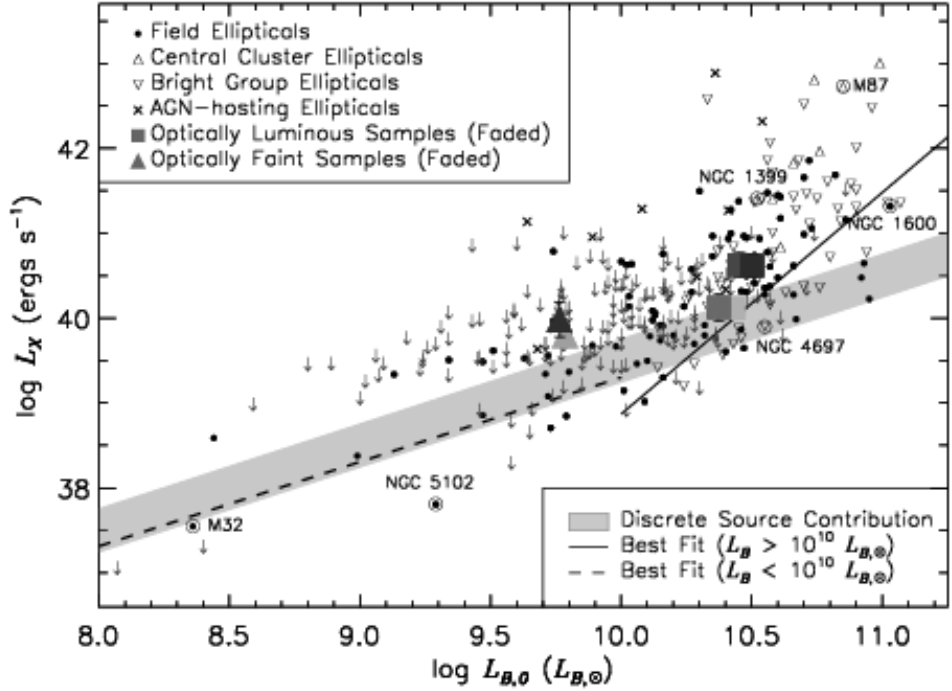}
}
\caption{
\footnotesize
Logarithm of the \hbox{0.5--2.0~keV} luminosity (\Lx) versus the
logarithm of the $B$-band luminosity, $L_B$, for $D < 70$~kpc local ellipticals with the
average properties of our faded samples plotted.  Small symbols and upper
limits are sources from the OS01 local sample, and the different symbols
correspond to field galaxies ({\it circles\/}), central-cluster galaxies ({\it
upward-pointing open triangles\/}), brightest-group galaxies ({\it
downward-pointing open triangles\/}), and AGNs ({\it crosses\/}).  The best-fit
relations for the luminosity intervals $L_B \simgt 10^{10}$~\lbsol\ and $L_B
\simlt 10^{10}$~\lbsol\ are shown as solid and dashed lines, respectively; the
shaded region shows the expected discrete-source contribution (Kim \& Fabbiano
2004; see also the discussion in $\S$~3.2.1).  The local ellipticals M32, M87,
NGC~1399, NGC~1600, NGC~4697, and NGC~5102 have been marked, for reference.
Large shaded symbols represent our optically luminous ({\it squares\/}) and
faint ({\it triangles\/}) faded samples and have grayscale levels corresponding
to the mean redshift of each sample, such that darker shading represents
higher redshifts.  Error bars in \Lx\ represent 1$\sigma$ errors on the mean.
We quote $B$-band luminosities as $L_{B,0}$ to illustrate any potential
evolution of the mean \xray\ luminosities.} \end{figure}

%
%

\begin{figure}
\figurenum{9}
\centerline{
\includegraphics[width=14cm]{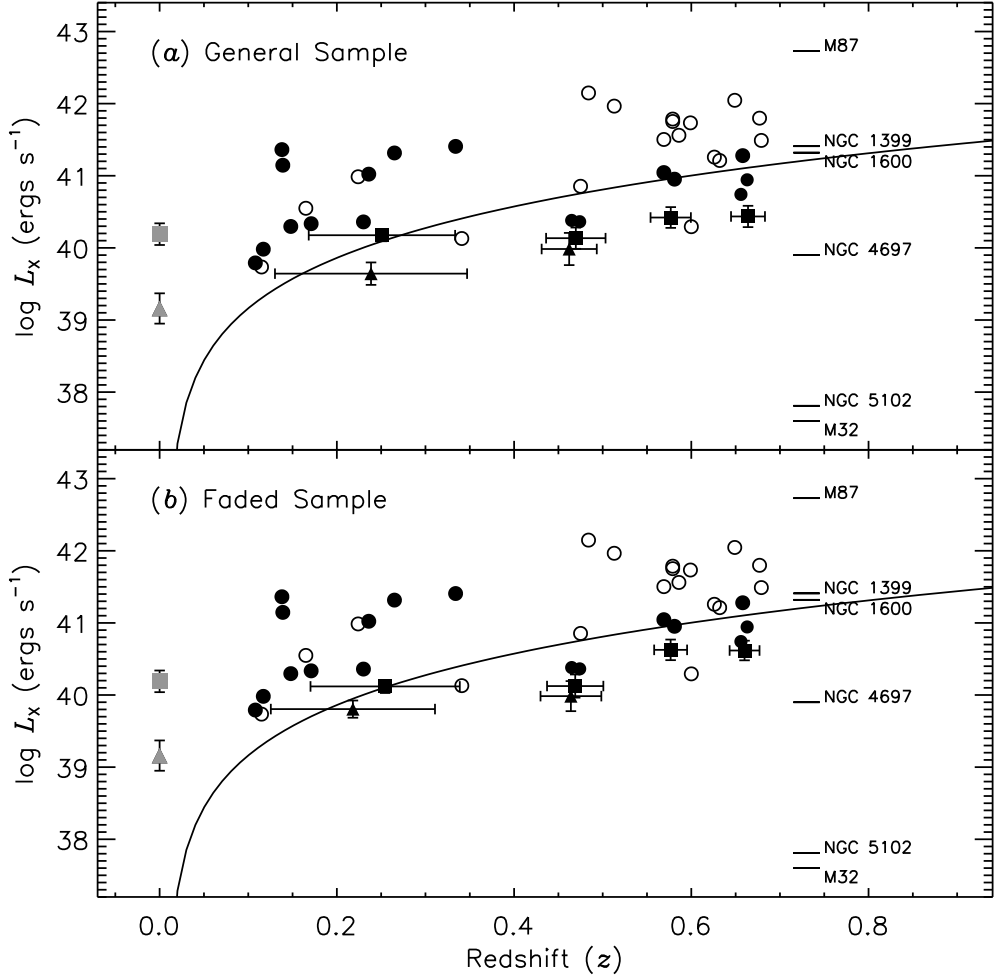}
}
\caption{
\footnotesize
Logarithm of the \hbox{0.5--2.0~keV} luminosity \Lx\ versus redshift for
our general ({\bf a}; see Figure~2a) and faded ({\bf b}; see Figure~2b)
samples; filled and open circles indicate \xray-detected normal galaxies and
AGN candidates, respectively.  Black squares and triangles with error bars show
the stacking results for our optically luminous ($L_B \approx
10^{10-11}$~\xlum) and faint ($L_B \approx 10^{9.3-10}$~\xlum) samples,
respectively.  Error bars in redshift represent the standard deviation of the
redshift for sources in each stacked sample.  For comparison, we have
plotted the corresponding mean \xray\ luminosities (and errors on the means
computed with {\ttfamily ASURV}) of normal early-type galaxies from the OS01
local sample ({\it gray filled square and triangle\/}).  All mean values (ours
and those of OS01) were calculated after excluding AGNs, central-cluster
galaxies, and brightest-group galaxies, and should reflect the average
properties of isolated field early-type galaxies. The solid curve illustrates
the median \xray\ detection limit (using the median sensitivity limit of
\hbox{$\approx$$1.6 \times 10^{-16}$~\flux} for our total sample).  For
reference, the \xray\ luminosities of the local ellipticals M32, M87, NGC~1399,
NGC~1600, NGC~4697, and NGC~5102 have been indicated.} \end{figure}

%
%

\begin{figure}
\figurenum{10}
\centerline{
\includegraphics[width=14cm]{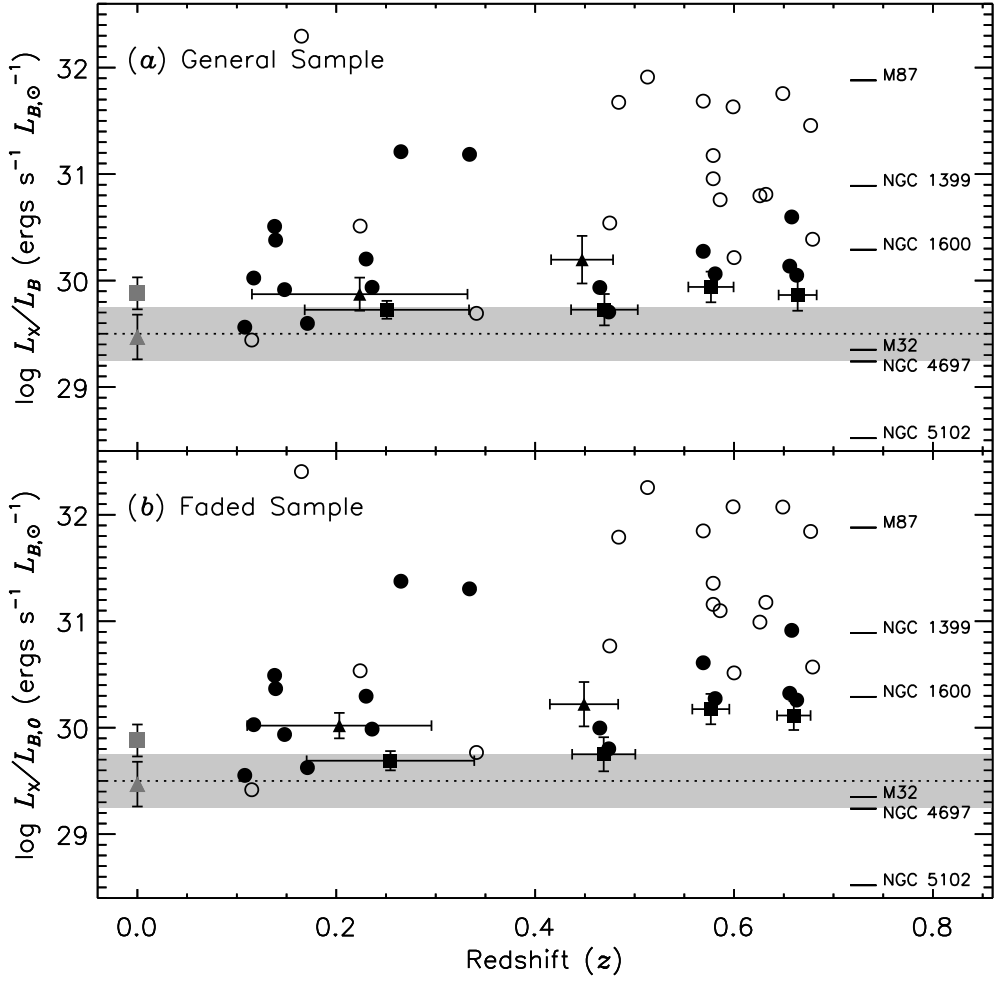}
}
\caption{
\footnotesize
Logarithm of the \xray--to--$B$-band mean luminosity ratio ($L_{\rm
X}/L_B$) versus redshift for our general ({\bf a}; see Figure~2a) and faded
({\bf b}; see Figure~2b) samples.  Symbols have the same meaning as in
Figure~9; the dotted line and shaded region represent the expected
local discrete-source contribution and its dispersion (from Kim \& Fabbiano
2004).  For the faded sample, we used $L_{B,0}$ when computing the
\xray--to--$B$-band mean luminosity ratios.  For reference, the local
ellipticals M32, M87, NGC~1399, NGC~1600, NGC~4697, and NGC~5102 have been
plotted.} \end{figure}

%
%

\begin{figure}
\figurenum{11}
\centerline{
\includegraphics[width=14cm]{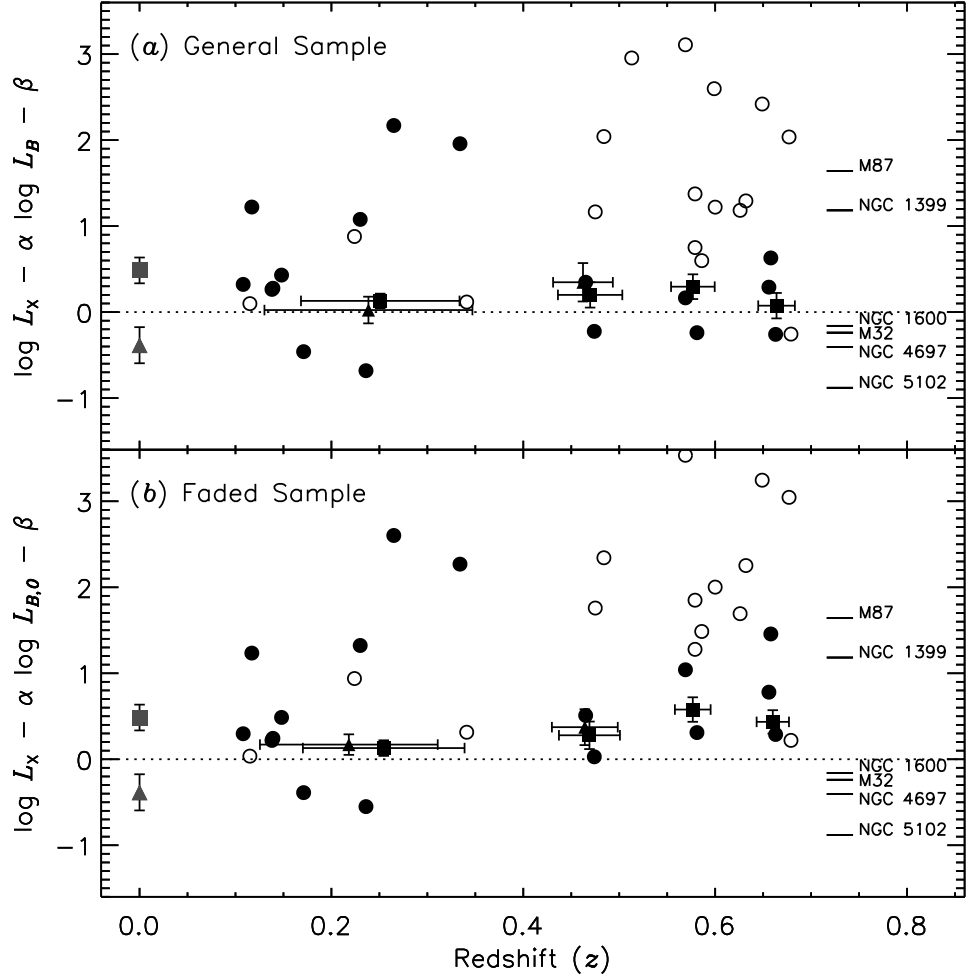}
}
\caption{
\footnotesize
Residuals to the local best-fit relation, $\log L_{\rm X} - \alpha \log
L_B - \beta$ for our general ({\bf a}; see Figure~2a) and faded ({\bf b}; see
Figure~2b) samples.  For optically luminous galaxies ($L_B \approx
10^{10-11}$~\lbsol) we used $\alpha=2.61$ and $\beta=12.77$, and for optically
faint galaxies ($L_B \approx 10^{9.3-10}$~\lbsol), we adopted $\alpha=1.05$ and
$\beta=29.36$.  Symbols have the same meaning as in Figure~9, and the redshifts
of the local samples have been offset from $z=0$ for viewing ease. The dotted
horizontal line indicates the zero residual.  For reference, the local
ellipticals M32, M87, NGC~1399, NGC~1600, NGC~4697, and NGC~5102 have been
plotted.} \end{figure}

%
%

\begin{figure}
\figurenum{12}
\centerline{
\includegraphics[width=10cm]{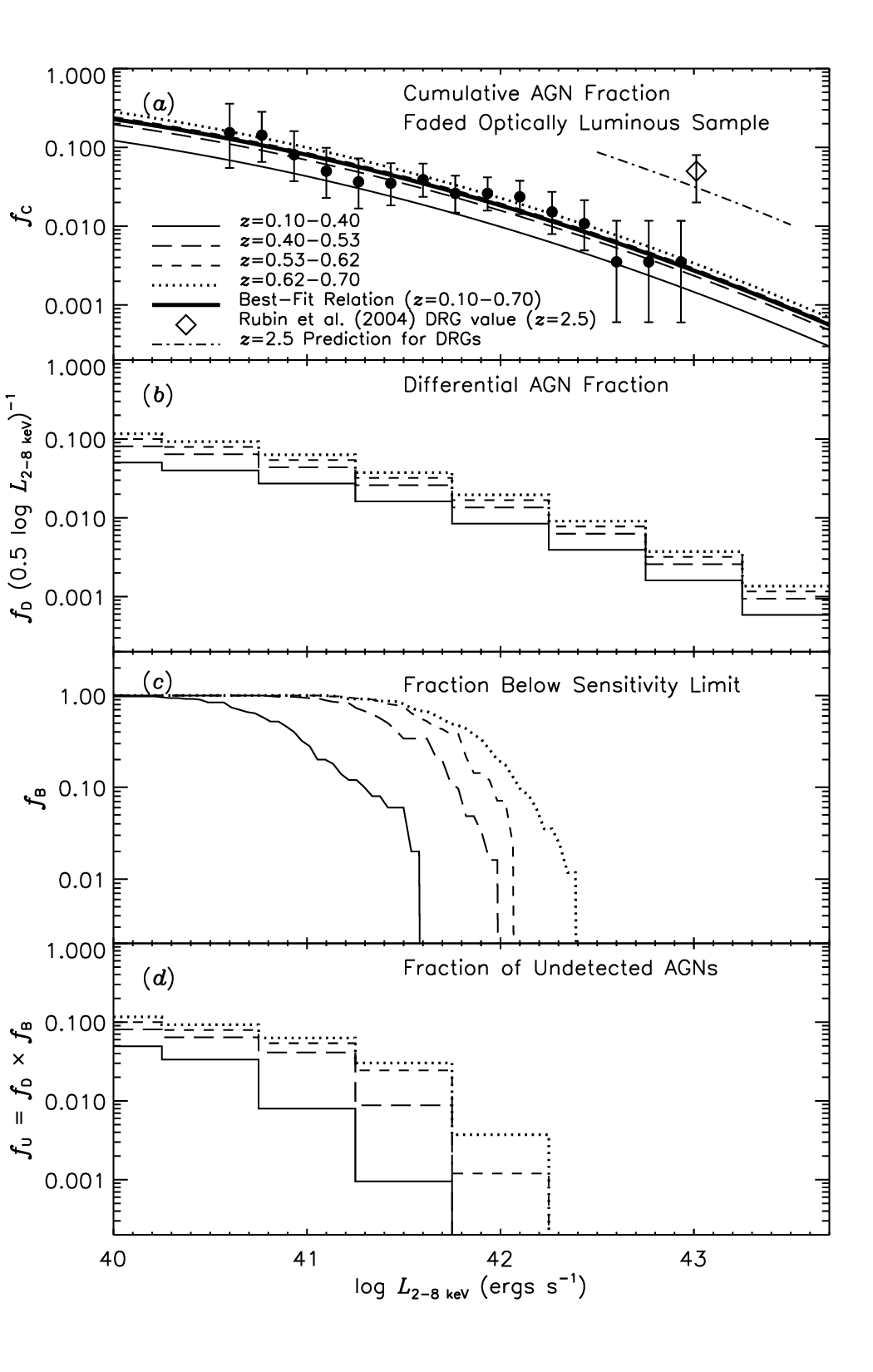}
}
\vspace{-0.3in}
\caption{
\footnotesize
({\bf a}) Cumulative AGN fraction (i.e., the fraction of galaxies
harboring an AGN with a \hbox{2--8~keV} luminosity of $L_{\rm 2-8~keV}$ or
greater), $f_C$, versus $\log L_{\rm 2-8~keV}$ for our optically luminous faded
samples.  The cumulative AGN fraction computed over the entire redshift range,
$\langle f_C \rangle_z$, is indicated as filled circles with 1$\sigma$ error bars, and the
thick solid curve represents our best-fit quadratic relation to the data (see
$\S$~3.2.2 for details).  This relation was used to estimate $f_C$ for our
optically luminous faded samples ({\it annotated curves\/}).  For reference, we
have plotted the observed AGN fraction for $z \approx 2.5$ DRGs (Rubin \etal
2004; {\it open diamond\/}) and our model extrapolated out to $z=2.5$ ({\it
dot-dashed curve\/}).  ({\bf b}) Differential AGN fractions (i.e., the fraction
of galaxies harboring an AGN in discrete bins of width $\Delta \log L_{\rm
2-8~keV}=0.5$), $f_D$, versus $\log L_{\rm 2-8~keV}$.  ({\bf c}) Fraction of
early-type galaxies for which we could {\it not} have detected an AGN with a
\hbox{2--8~keV} luminosity of $L_{\rm 2-8~keV}$, $f_B$, versus $\log L_{\rm
2-8~keV}$.  ({\bf d}) Fraction of galaxies harboring AGNs in our optically
luminous faded samples that would remain undetected due to sensitivity
limitations, $f_U = f_D \times f_B$, versus $\log L_{\rm 2-8~keV}$; these
galaxies would not have been removed from our stacking analyses.  }
\end{figure}

%
%

\begin{figure}
\figurenum{13}
\centerline{
\includegraphics[width=14cm]{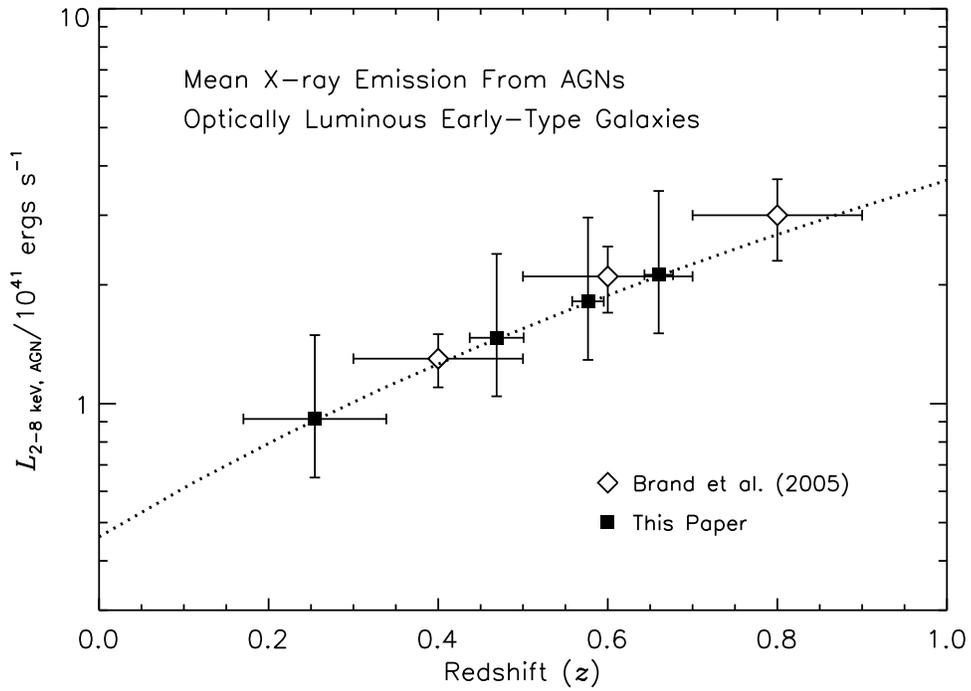}
}
\caption{
\footnotesize
Mean \hbox{2--8~keV} AGN luminosity per early-type galaxy, $L_{\rm 2-8~keV,
AGN}$, versus redshift for our optically luminous faded samples ({\it filled
squares\/}); these luminosities were derived following equation~8.  Error bars
on $L_{\rm 2-8~keV, AGN}$ are 1$\sigma$ errors, which were derived by propogating
errors on the fit to $\langle f_C \rangle_z$ ({\it solid curve} in Fig.~12)
through to equation~8.  For comparison, we have plotted the Brand \etal (2005)
stacking constraints for their early-type galaxy samples ({\it open
diamonds\/}), which have mean $R$-band absolute magnitudes similar to those of
our optically luminous faded samples.  The dotted curve represents the
$(1+z)^3$ model fit to our data.} \end{figure}

\end{document}